\documentclass[longauth]{aa}
\usepackage{graphicx}
\usepackage[section]{placeins}
\usepackage{sidecap}
\usepackage[english]{babel}
\usepackage{txfonts}
\usepackage{multirow}
\usepackage{natbib}
\usepackage{color}
\usepackage{url}
\bibpunct{(}{)}{;}{a}{}{,} 
\usepackage{amssymb}

\newcommand{\OII}{$\left[\mathrm{O\textrm{\textsc{ii}}}\right]\,$}

\newcommand{\OIII}{$\left[\mathrm{O\textrm{\textsc{iii}}}\right]\,$}

\newcommand{\Ha}{H${\alpha}\,$}
\newcommand{\Hb}{H${\beta}\,$}

\newcommand{\uflux}[0]{${\rm erg \cdot s^{-1} \cdot cm^{-2}}$}

\begin{document}

\title{SDSS-IV eBOSS emission-line galaxy pilot survey} 
\author{J. Comparat,\inst{1,2,}
\thanks{j.comparat@csic.es}$^,$\thanks{Severo Ochoa IFT Fellow} \and 
T. Delubac\inst{3} \and 
S. Jouvel\inst{4} \and 
A. Raichoor\inst{5} \and 
J-P. Kneib\inst{3,6} \and 
C. Y\`eche\inst{5} \and 
F. B. Abdalla\inst{4,7} \and 
C. Le Cras\inst{8} \and 
C. Maraston\inst{8} \and 
D. M. Wilkinson\inst{8} \and 
G. Zhu\inst{9} \and 
E. Jullo\inst{6} \and 
F. Prada\inst{1,2,10} \and 
D. Schlegel\inst{11} \and 
Z. Xu\inst{12} \and H. Zou\inst{12} \and 
J. Bautista\inst{13} \and 
D. Bizyaev\inst{14,15} \and 
A. Bolton\inst{13} \and 
J. R. Brownstein\inst{13} \and 
K. S. Dawson\inst{13} \and 
S. Escoffier\inst{16}  \and 
P. Gaulme\inst{14} \and 
K. Kinemuchi\inst{14} \and 
E. Malanushenko\inst{14} \and 
V. Malanushenko\inst{14} \and 
V. Mariappan\inst{13} \and 
J. A. Newman\inst{17} \and 
D. Oravetz\inst{14} \and 
K. Pan\inst{14} \and 
W. J. Percival\inst{8} \and
A. Prakash\inst{17} \and
D. P. Schneider\inst{18,19} \and
A. Simmons\inst{14} \and
T. M. C.~Abbott\inst{20}
S.~Allam\inst{21} \and
M.~Banerji\inst{22,23} \and
A.~Benoit-L{\'e}vy\inst{4} \and
E.~Bertin\inst{24,25} \and
D.~Brooks\inst{4} \and
D.~Capozzi\inst{8} \and
A.~Carnero~Rosell\inst{26,27} \and
M.~Carrasco~Kind\inst{28,29} \and
J.~Carretero\inst{30,31} \and
F.~J.~Castander\inst{31} \and
C.~E.~Cunha\inst{32} \and
L.~N.~da Costa\inst{26,27} \and
S.~Desai\inst{33,34} \and
P.~Doel\inst{4} \and
T.~F.~Eifler\inst{35,36} \and
J.~Estrada\inst{21} \and
B.~Flaugher\inst{21} \and
P.~Fosalba\inst{30} \and
J.~Frieman\inst{21,37} \and
E.~Gaztanaga\inst{30} \and
D.~W.~Gerdes\inst{38} \and
D.~Gruen\inst{39,40} \and
R.~A.~Gruendl\inst{28,29} \and
G.~Gutierrez\inst{21} \and
K.~Honscheid\inst{41,42} \and
D.~J.~James\inst{20} \and
K.~Kuehn\inst{43} \and
N.~Kuropatkin\inst{21} \and
O.~Lahav\inst{4} \and
M.~Lima\inst{44,26} \and
M.~A.~G.~Maia\inst{26,27} \and
M.~March\inst{35} \and
J.~L.~Marshall\inst{45} \and
R.~Miquel\inst{46,31} \and
A.~A.~Plazas\inst{36} \and
K.~Reil\inst{47} \and
N.~Roe\inst{11} \and
A.~K.~Romer\inst{48} \and
A.~Roodman\inst{32,47} \and
E.~S.~Rykoff\inst{32,47} \and
M.~Sako\inst{35} \and
E.~Sanchez\inst{49} \and
V.~Scarpine\inst{21} \and
I.~Sevilla-Noarbe\inst{49,28} \and
M.~Soares-Santos\inst{21} \and
F.~Sobreira\inst{21,26} \and
E.~Suchyta\inst{41,42} \and
M.~E.~C.~Swanson\inst{29} \and
G.~Tarle\inst{38} \and
J.~Thaler\inst{50} \and
D.~Thomas\inst{8} \and
A.~R.~Walker\inst{20} \and
Y.~Zhang\inst{38}}

\institute{
Instituto de Fisica Teorica  UAM/CSIC, Universidad Aut\'onoma de Madrid, Cantoblanco, E-28049 Madrid, Spain\\ \and 
Departamento de Fisica Teorica, Universidad Aut\'onoma de Madrid, Cantoblanco, E-28049 Madrid,  Spain\\ \and 
Laboratoire d'Astrophysique, Ecole Polytechnique F\'ed\'erale de Lausanne (EPFL), Observatoire de Sauverny, CH-1290 Versoix, Switzerland\\ \and 
Department of Physics and Astronomy, University College London, Gower Street, London WC1E6BT, UK\\ \and 
CEA, Centre de Saclay, IRFU/SPP, F-91191 Gif-sur-Yvette, France\\ \and 
Aix Marseille Universit\'e, CNRS, LAM (Laboratoire d'Astrophysique de Marseille) UMR 7326, F-13388, Marseille, France\\ \and 
Department of Physics \& Electronics, Rhodes University, Grahamstown 6140, South Africa\\ \and 
Institute of Cosmology and Gravitation, University of Portsmouth, Portsmouth, PO1 3FX, UK\\ \and 
Department of Physics \& Astronomy, Johns Hopkins University, 3400 N. Charles Street, Baltimore, MD 21218, USA\\ \and 
Instituto de Astrof\'isica de Andaluc\'ia (CSIC), Glorieta de la Astronom\'ia, E-18080 Granada, Spain\\ \and 
Lawrence Berkeley National Laboratory, 1 Cyclotron Road, Berkeley, CA, 94720, USA\\ \and 
Key Laboratory of Optical Astronomy, National Astronomical Observatories, Chinese Academy of Sciences, Beijing, 100012,China \\ \and 
Department of Physics and Astronomy, University of Utah, 115 S 1400 E, Salt Lake City, UT 84112, USA\\ \and 
Apache Point Observatory and New Mexico State University, P.O. Box 59, Sunspot, NM, 88349-0059, USA \\ \and 
Sternberg Astronomical Institute, Moscow State University, Moscow \\ \and 
CPPM, Aix-Marseille Universit\'e, CNRS/IN2P3, Marseille, France\\ \and 
Department of Physics and Astronomy and PITT PACC, University of Pittsburgh, Pittsburgh, PA 15260, USA\\ \and 
Department of Astronomy and Astrophysics, The Pennsylvania State University, University Park, PA 16802\\ \and 
Institute for Gravitation and the Cosmos, The Pennsylvania State University, University Park, PA 16802 \\ \and 
Cerro Tololo Inter-American Observatory, National Optical Astronomy Observatory, Casilla 603, La Serena, Chile\\ \and 
Fermi National Accelerator Laboratory, P. O. Box 500, Batavia, IL 60510, USA\\ \and 
Institute of Astronomy, University of Cambridge, Madingley Road, Cambridge CB3 0HA, UK\\ \and 
Kavli Institute for Cosmology, University of Cambridge, Madingley Road, Cambridge CB3 0HA, UK\\ \and 
CNRS, UMR 7095, Institut d'Astrophysique de Paris, F-75014, Paris, France\\ \and 
Sorbonne Universit\'es, UPMC Univ Paris 06, UMR 7095, Institut d'Astrophysique de Paris, F-75014, Paris, France\\ \and 
Laborat\'orio Interinstitucional de e-Astronomia - LIneA, Rua Gal. Jos\'e Cristino 77, Rio de Janeiro, RJ - 20921-400, Brazil\\ \and 
Observat\'orio Nacional, Rua Gal. Jos\'e Cristino 77, Rio de Janeiro, RJ - 20921-400, Brazil\\ \and 
Department of Astronomy, University of Illinois, 1002 W. Green Street, Urbana, IL 61801, USA\\ \and 
National Center for Supercomputing Applications, 1205 West Clark St., Urbana, IL 61801, USA\\ \and 
Institut de Ci\`encies de l'Espai, IEEC-CSIC, Campus UAB, Carrer de Can Magrans, s/n,  08193 Bellaterra, Barcelona, Spain\\ \and 
Institut de F\'{\i}sica d'Altes Energies, Universitat Aut\`onoma de Barcelona, E-08193 Bellaterra, Barcelona, Spain\\ \and 
Kavli Institute for Particle Astrophysics \& Cosmology, P. O. Box 2450, Stanford University, Stanford, CA 94305, USA\\ \and 
Excellence Cluster Universe, Boltzmannstr.\ 2, 85748 Garching, Germany\\ \and 
Faculty of Physics, Ludwig-Maximilians University, Scheinerstr. 1, 81679 Munich, Germany\\ \and 
Department of Physics and Astronomy, University of Pennsylvania, Philadelphia, PA 19104, USA\\ \and 
Jet Propulsion Laboratory, California Institute of Technology, 4800 Oak Grove Dr., Pasadena, CA 91109, USA\\ \and 
Kavli Institute for Cosmological Physics, University of Chicago, Chicago, IL 60637, USA\\ \and 
Department of Physics, University of Michigan, Ann Arbor, MI 48109, USA\\ \and 
Max Planck Institute for Extraterrestrial Physics, Giessenbachstrasse, 85748 Garching, Germany\\ \and 
Universit\"ats-Sternwarte, Fakult\"at f\"ur Physik, Ludwig-Maximilians Universit\"at M\"unchen, Scheinerstr. 1, 81679 M\"unchen, Germany\\ \and 
Center for Cosmology and Astro-Particle Physics, The Ohio State University, Columbus, OH 43210, USA\\ \and 
Department of Physics, The Ohio State University, Columbus, OH 43210, USA\\ \and 
Australian Astronomical Observatory, North Ryde, NSW 2113, Australia\\ \and 
Departamento de F\'{\i}sica Matem\'atica,  Instituto de F\'{\i}sica, Universidade de S\~ao Paulo,  CP 66318, CEP 05314-970, S\~ao Paulo, SP,  Brazil\\ \and 
George P. and Cynthia Woods Mitchell Institute for Fundamental Physics and Astronomy, and Department of Physics and Astronomy, Texas A\&M University, College Station, TX 77843,  USA\\ \and 
Instituci\'o Catalana de Recerca i Estudis Avan\c{c}ats, E-08010 Barcelona, Spain\\ \and 
SLAC National Accelerator Laboratory, Menlo Park, CA 94025, USA\\ \and 
Department of Physics and Astronomy, Pevensey Building, University of Sussex, Brighton, BN1 9QH, UK\\ \and 
Centro de Investigaciones Energ\'eticas, Medioambientales y Tecnol\'ogicas (CIEMAT), Madrid, Spain\\ \and 
Department of Physics, University of Illinois, 1110 W. Green St., Urbana, IL 61801, USA"\\
}

\date{\today}

\abstract
{The Sloan Digital Sky Survey IV extended Baryonic Oscillation Spectroscopic Survey (SDSS-IV/eBOSS) will observe 195,000 emission-line galaxies (ELGs) to measure the Baryonic Acoustic Oscillation standard ruler (BAO) at redshift 0.9.
To test different ELG selection algorithms, 9,000 spectra were observed with the SDSS spectrograph as a pilot survey based on data from
several imaging surveys. First, using visual inspection and redshift quality flags, we show that the automated spectroscopic redshifts assigned by the pipeline meet the quality requirements for a reliable BAO measurement. We also show the correlations between sky emission, signal-to-noise ratio in the emission lines, and redshift error. Then we provide a detailed description of each target selection algorithm we tested and compare them with the requirements of the eBOSS experiment. 
As a result, we provide reliable redshift distributions for the different target selection schemes we tested.
Finally, we determine an target selection algorithms that
is best suited to be applied on DECam photometry because they fulfill the eBOSS survey efficiency requirements.
}
\titlerunning{eBOSS ELG}

\maketitle
\section{Introduction}
\label{sec:introduction}

Galaxy surveys permit studying the cosmological structures formed by the network of galaxies and the evolution of galaxies.
The recent increase in the number of multiplexing of spectrographs \citep{2003SPIE.4841.1670L,2006SPIE.6269E..0GS,2013AJ....146...32S,2014SPIE.9147E..0SF} 
and in the field of view of photometric cameras \citep{1998AJ....116.3040G,2003SPIE.4841...72B,2015arXiv150402900F} 
allows galaxy surveys to cover larger areas of the sky and to measure large numbers of accurate redshifts.
The precision of a cosmological statement based on a galaxy survey is directly 
related to the volume sampled by the survey: the larger, the better.
To extract cosmological information from a magnitude-limited
 galaxy survey, 
we therefore construct the largest possible volume-limited sample 
\citep{2002SPIE.4847...86M}.
To enhance the covered volume and increase survey efficiency, galaxy surveys push the high-redshift limit as far as possible and pre-select galaxies by discarding low-redshift faint galaxies to obtain an observed sample as close as possible to the desired volume-limited sample. 

Three recent surveys successfully applied a color selection to a magnitude-limited sample to map a given redshift range more efficiently and to extract cosmological information. The Baryonic Oscillation Spectroscopic Survey \citep[BOSS: ][]{2013AJ....145...10D,2011AJ....142...72E} observed a specific part of the galaxy population, the most massive ellipticals at redshift 0.57, by selecting in the color-color diagram $g-r$, $r-i$ \citep{2013MNRAS.435.2764M}. BOSS is the first to measure the Baryonic Acoustic Oscillation standard ruler at the percent level\citep[BAO:][]{2003ApJ...598..720S}, which
directly constrains the cosmological model \citep{2014MNRAS.441...24A}. The WiggleZ survey targeted star-forming galaxies at redshift 0.6 \citep{2010MNRAS.401.1429D} and measured the BAO standard ruler at the 5 \% level \citep{2014MNRAS.441.3524K}. Finally, the VIMOS Public Extragalactic Redshift Survey \citep[VIPERS,][]{2014A&A...566A.108G} observed galaxies at redshift 0.8 and measured the growth rate of structure at the 17\% level \citep{2013A&A...557A..54D}. 

Measuring standard rulers and the growth rate of structure is key to understand the cosmological model \citep{2013PhR...530...87W}, but this does not require the galaxy sample to be volume limited.
As illustrated by the measurement of the BAO feature using samples that are not volume limited, such as the Ly$\alpha$ forest of quasars \citep{2015A&A...574A..59D} or the WiggleZ survey, the BAO feature is an intrinsic property of the matter field and does not require a volume-limited sample. However, it is mandatory to understand in depth how the selected tracers are related to the dark matter field \citep{2015JCAP...05..060B,2015MNRAS.449.1454P,2015arXiv150704356F,2015arXiv150906404R}. Furthermore, \citet{2014MNRAS.442.2131A} found that galaxy formation effects, such as the selection function or bias models, can bias the BAO scale measurement at the 0.2\% level, which is well below the expected precision by eBOSS.
Therefore, we can spectroscopically observe any tracer of the matter field. In particular, we can choose any type of galaxy that is believed to be a Poisson sampling of the underlying density field, provided its density is sufficient to overcome the shot noise and that accurate redshifts can be obtained in a short time, so that a large ($>h^{-3}$Gpc$^3$) volume can be covered within a few years of observations.

The extended Baryonic Oscillation Spectroscopic Survey \citep[eBOSS,][]{2015arXiv150804473D} uses color selection to identify three types of galaxies: luminous red galaxies \citep[LRGs,][]{2015arXiv150804478P}, emission-line galaxies (ELGs), and quasars \citep[QSOs,][]{2015ApJS..221...27M}. The ELG and QSO samples will not provide volume-limited samples. 
 eBOSS will measure the angle-averaged BAO distance ladder with point-like tracers to the 1\% precision at redshift 0.7 using LRGs , 2.2\% at redshift 0.87 using ELGs (or d$_A$(z) to an accuracy of 3.1\% and H(z) to 4.7\%), and 1.6\% at redshift 1.37 using QSOs \citep{2015arXiv150804473D,2016MNRAS.457.2377Z}. We expect to reach this measurement by acquiring 195,000 emission line galaxy redshifts at an effective redshift of z = 0.87 using 300 dedicated plates. The efficiency of the targeting algorithm needs to be higher than $N(0.6<z_\mathrm{reliable}<1)/N_\mathrm{targets}>74.5\%$ or $N(0.7<z_\mathrm{reliable}<1.1)/N_\mathrm{targets}>74.5\%$ to obtain the correct amount of reliable redshifts in the range of interest. Additionally, the fraction of catastrophic redshifts in the range $0.6<z<1$ or $0.7<z<1.1$ must be below 1\%. 

The target selection of the ELG for BAO surveys is driven by the requirement to acquire as many spectra as possible in the smallest amount of observing time to maximize the volume covered by the survey. 
The resulting precision on the measurement of the BAO scale is directly related to the effective volume sample, that is, the volume where the density of tracers is above shot noise. For a constant density as a function of redshift, increasing the area is the only handle to increase the effective volume. For a peaked redshift distribution, increasing the density also increases the effective volume by overcoming shot noise at the tails of the redshift distribution. 
The trade-off of density vs. area was studied, and we found that targeting 340 ELG deg$^{-2}$ on 750 deg$^2$ or 170 ELG deg$^{-2}$ on 1,500 deg$^2$ would provide the same effective volume and therefore the same BAO measurement \citep{2015arXiv150804473D}. 
We present here the pilot data that were observed to investigate the possible target selection algorithms using photometric data sets  that are available on an area at least larger than 750 deg$^2$ and that can provide a density of targets, denoted $\rho$, such as $170<\rho<340$ deg$^{-2}$. These target selection algorithms must fulfill the redshift efficiency requirement mentioned above. 

For both ELGs and QSOs, schemes that select targets for spectroscopy exist using either color selection \citep{2013MNRAS.428.1498C,2010AJ....139.2360S} or higher dimensional algorithms \citep{2016A&A...585A..50R,2012ApJ...749...41B}. 
To assess algorithms, eBOSS tested them on a 10 deg$^2$ sky patch covered by many photometric surveys around $\alpha\sim36^\circ$ and $\delta\sim-4.5^\circ$. This region was indeed observed by the following surveys SDSS, CFHT-LS Wide, DES, SCUSS, and WISE \citep[][respectively]{2012ApJS..203...21A,2012AJ....143...38G,2015MNRAS.446.2523B,1538-3881-150-4-104,2010AJ....140.1868W}.

In addition to finding the right algorithm, we face another challenge:
we need to develop the best targeting algorithm implementable with existing photometry to construct a homogeneous ELG sample. 

This paper is included in a series of eBOSS papers. \citet{2015arXiv150804473D} gave an overview of the eBOSS survey. \citet{2016MNRAS.457.2377Z} provided the Fisher matrix forecast on the accuracy of the BAO and RSD measurements expected. This paper presents the results of the ELG pilot survey that enables a complete study of the ELG target selection algorithm. We mitigated the risk by investigating TS algorithms using existing well-known photometry (SDSS and WISE) and more recent deeper ones (DES and SCUSS). \citet{2016A&A...585A..50R} described in depth the optimization of the SDSS+WISE+SCUSS algorithms using the Fisher technique, while the further optimization of algorithms using DES photometry is described in this paper. The clustering properties and the homogeneity of targeting catalogs on large areas are discussed in \citet{2015arXiv150907121J} for the DES-based TS and in Delubac et al. (in prep) for the SDSS+WISE -based TS.

In this paper, we analyze the pilot survey observations from eBOSS ELG carried out at the SDSS telescope \citep{2006AJ....131.2332G}. 
In Sect. \ref{sec:DATA} we describe the photometric catalog from which the targets were drawn and how the spectroscopy of the ELG was performed.
Section \ref{sec:BOSS:ELG:redshifts} describes how the redshifts are automatically measured. 
Section \ref{sec:ELGTS:algorithms} details the exact selection algorithms applied and the corresponding galaxy population observed. 
In Sect. \ref{sec:decamTS} we introduce an optimized selection scheme based on DECAM imaging that is suited for the eBOSS ELG observations.

Throughout the paper, we quote magnitudes in the AB system \citep{1983ApJ...266..713O} and provide the measurements in a flat $\Lambda$CDM cosmology $h=0.7$, $\Omega_m=0.3$.

\section{Data}
\label{sec:DATA}

\subsection{Photometry}

To select targets, we used photometry from the following surveys.

\subsubsection*{SDSS, WISE, and SCUSS}
The SDSS photometry \citep{2015ApJS..219...12A}\footnote{\url{http://www.sdss.org/dr12/imaging/}} is constituted of the five broad bands $u$, $g$, $r$, $i$, $z$ \citep{1996AJ....111.1748F} and covers about 14,000 deg$^2$.

We computed so-called forced photometry on the SDSS $r$-band detected objects, which is 75\% complete at 22.5  \citep{2012ApJS..203...21A,2014arXiv1410.7397L}. 
It uses measured SDSS source positions, star-galaxy separation, and galaxy profiles to define the sources whose fluxes are to be measured in the WISE images \citep{2010AJ....140.1868W}. The WISE imaging has a 6.1 arc seconds seeing in the band W1 at 3.4$\mu m$.

We used data from the SCUSS survey, which is a u-band survey with a magnitude limit of 23.2 with a seeing of $\sim$2 arc seconds \citep{1538-3881-150-4-104}. We also apply the SDSS-based forced photometry model to these data \citep{2015PASP..127...94Z}. It constructs 2D models (de Vaucouleurs and exponential) based on SDSS r-band galaxy profiles and star-galaxy separation, and estimates object fluxes through comparing the models with the object images of SCUSS. The modelMag magnitudes in SCUSS are derived from the object flux with higher likelihood in the de Vaucouleurs and exponential model fitting.

Problems related to colors computed with magnitudes measured through different surveys are significantly mitigated for our SDSS-SCUSS-WISE colors because our SCUSS and WISE photometry is acquired consistently with the SDSS (forced photometry on SDSS objects, using SDSS structural information: \citealt{2014arXiv1410.7397L,2015PASP..127...94Z}).

We created a multiband catalog of detections. The SDSS-SCUSS-WISE catalog extends the full South Galactic Cap; see \citet{2016A&A...585A..50R} and Delubac et al. (in prep.) for the complete description of the catalog.

\subsubsection*{DES and DECaLS}
The Dark Energy Camera \citep[DECam][]{2015arXiv150402900F} was mounted, installed, and commissioned in 2012 on the Blanco four-meter telescope at Cerro Tololo Inter-American Observatory in Chile \citep{2012SPIE.8446E..6FD}. 

The Dark Energy Survey (DES) is an imaging survey of the grizY photometric bands \citep{2013AAS...22133502F,2014SPIE.9149E..0VD} that will cover 5000 deg$^2$ to an unprecedented 5$\sigma$ depth of $i_{AB}<24$ \citep{2012APS..APR.D7007F} using DECam.
A Science Verification (SV) period followed, and data were acquired between end of 2012 through February 2013. As the data were taken shortly after DECam commissioning and 
were used to test survey operations and assess data quality, the 
DES-SV data quality is not as good as the full survey data is 
expected to be, but it was deep enough for our purpose.
The SV fields were chosen to cover sky areas observed by spectroscopic galaxy surveys such as DEEP2 \citep{2013ApJS..208....5N} and VVDS \citep{2013A&A...559A..14L}. To test the target selections, we used the DES-SV around the CFHTLS-W1 field. 

The DECaLS photometry is part of the {\it Legacy Survey}\footnote{\url{http://legacysurvey.org/}}, which is producing an inference model catalog of the sky from a set of optical and infrared imaging data, comprising 14,000 deg$^2$ of extragalactic sky visible from the northern hemisphere in three optical bands (g,r,z) and four infrared bands (w1, w2, w3, w4 from WISE). The sky coverage that will eventually be surveyed by DECam is approximately bounded by $-18^\circ < \delta < +30^\circ$ in celestial coordinates and $|b| > 18^\circ$ in Galactic coordinates. This survey should provide a sufficient area to target ELGs from in a zone accessible to the SDSS telescope.

The DES and DECaLS photometries are about ten times deeper than the SDSS photometry and therefore allow a narrower and more efficient redshift selection; see the discussion in \citet{2013MNRAS.428.1498C}. For instance, because the scatter in the magnitude-color or color-color diagrams is smaller, it permits cleaner selections. Given that the area available to target from DECam-based imaging is smaller, we tried denser selections that extend to higher redshifts.

Finally, we created a matched detection catalog from the SDSS, CFHT-LS Wide, DES, SCUSS, and WISE photometric surveys. In the match all entries from the different catalogs are kept, even when they did not have a match in the other catalogs.

\subsection{Spectroscopy}
\begin{figure*}
\begin{center}
\includegraphics[width=\textwidth]{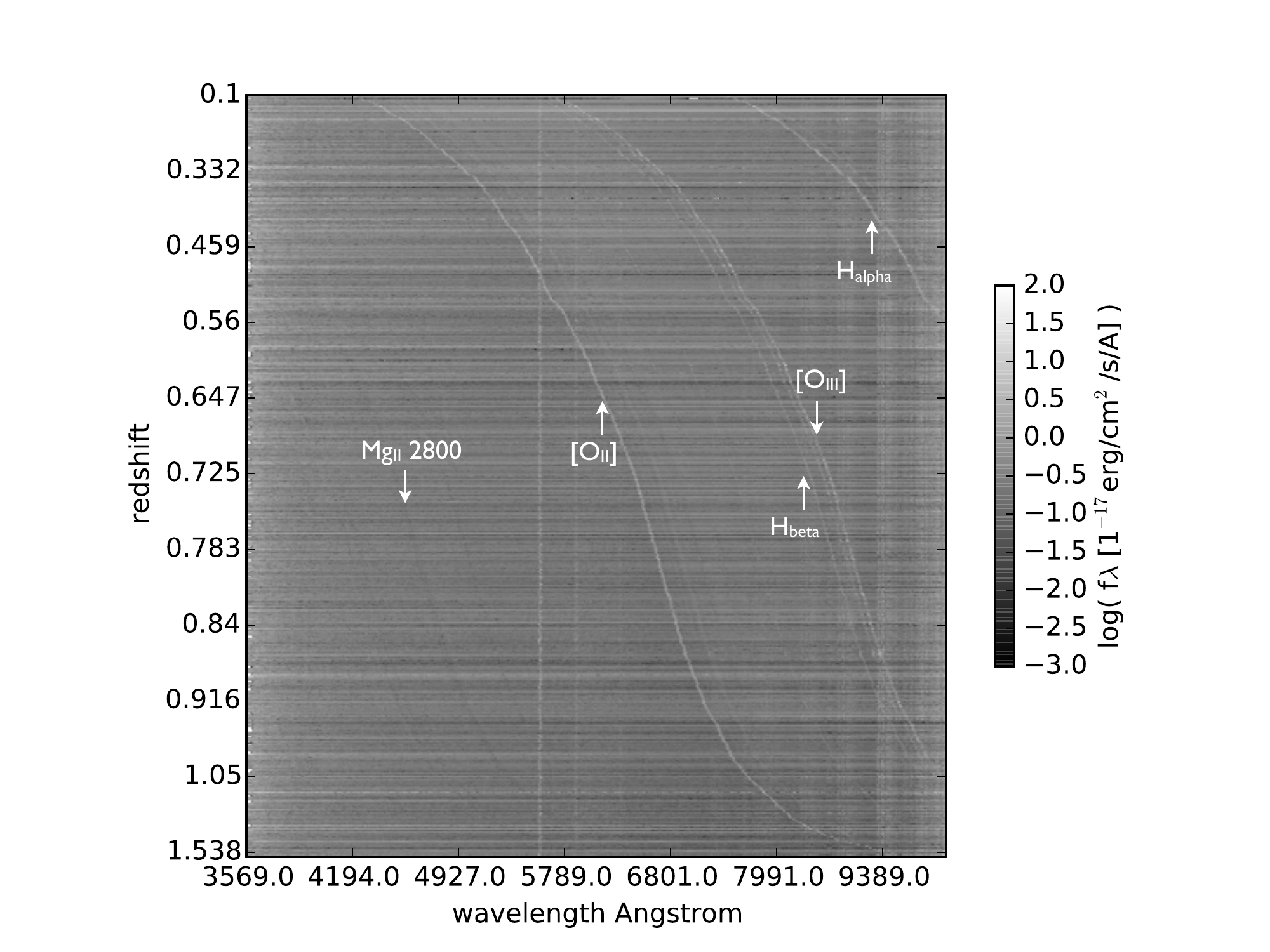}
\includegraphics[width=\textwidth]{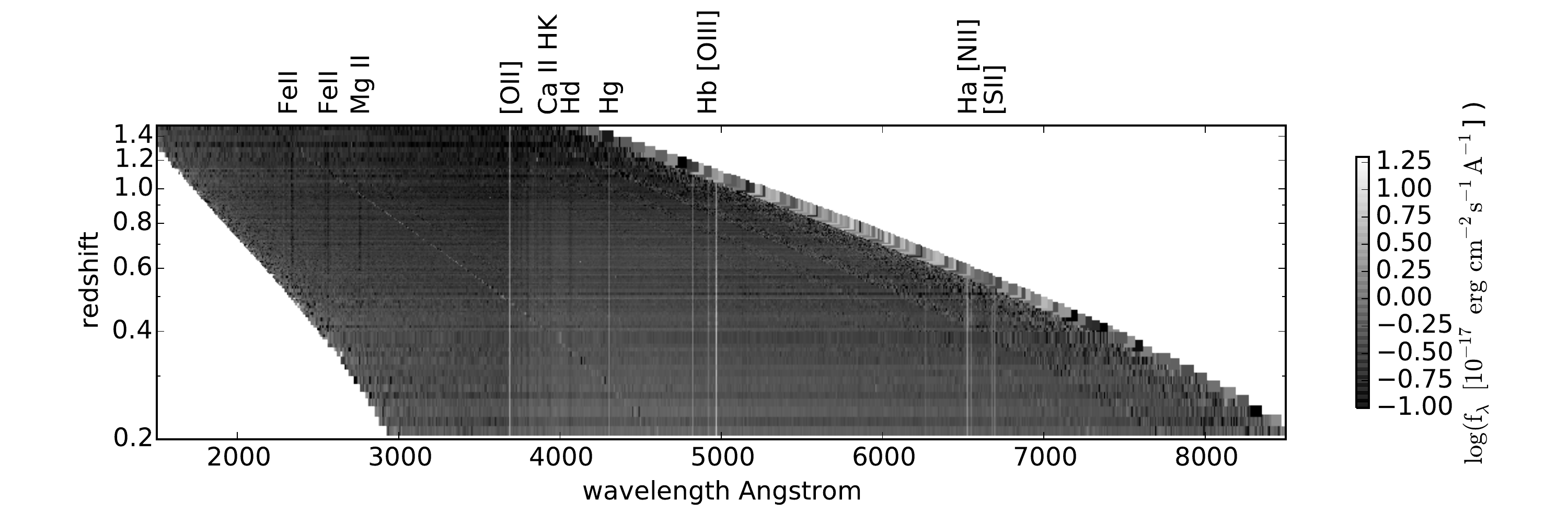}
\caption{\textbf{Top panel}. ELG spectra represented in the observed frame sorted by redshift. Each horizontal line corresponds to one spectrum. Vertical patterns correspond to the residuals of the sky subtraction. Diagonal patterns are the emission and absorption lines seen in the galaxy spectra. As redshift increases, the \OII, \Hb, \OIII, and \Ha emission lines become redshifted. Starting at redshift 0.5, UV absorption lines enter the spectrograph window. \textbf{Bottom panel} ELG spectra stacked by 50, ordered as a function of redshift represented in the rest-frame. The spectral features align vertically and are detected with higher signal-to-noise-ratio, in particular the absorption lines in the UV. Extended details about the UV absorption and emission systems are given in \citet{2015ApJ...815...48Z}.}
\label{fig:all:spectra}
\end{center}
\end{figure*}

The SDSS-BOSS spectrograph is an optical multifiber spectrograph mounted on the 2.5 m f/5 modified Ritchey-Chretien altitude-azimuth telescope located at the Apache Point Observatory, \citep[APO][]{2006AJ....131.2332G,2013AJ....146...32S}. SDSS spectra cover the wavelength range $\rm 3,600<\lambda<10,400\AA$ at an average resolution of 2000. The wavelength $\lambda$ is calibrated to vacuum wavelengths. The BOSS spectrophotometric calibration is accurate at the $<$5\% level in the $r$ band and $<$10\% in the other bands \citep{2015ApJS..216....4S}. 

\subsection{Previous ELG observations with the SDSS spectrograph}
During the SDSS-III BOSS survey, 11,883 ELG were observed and visually inspected \citep[see][]{2013MNRAS.428.1498C,2015A&A...575A..40C,2015ApJS..219...12A}. We added this sample to the eBOSS pilot survey to study the reliability of automated redshift assignment. The plates that contain ELGs are 4386-4389, 4391, 4392, 4394, 4395, 4397, and 4399 with a sparse ELG sampling and 5017, 5018, 6931-6933, 7239-7243, and 7245-7247 that are dedicated plates. The redshifts are estimated with the pipeline described in \citet{2012AJ....144..144B}.

\subsection{ELG pilot observations for SDSS-IV/eBOSS}
At the APO facility, 9000 fibers, or ten plates, were dedicated to the eBOSS ELG pilot survey. The exposure time was 4 $\times$ 15 minutes for each plate. 
The test extended over $\sim$10 deg$^2$ and was located around $\alpha$(J2000)$\sim36^\circ$ and $\delta$(J2000)$\sim-4.5^\circ$. They were labeled as `chunks' -- SDSS jargon for an observational run -- `eboss6' (plates 8123 to 8130) and `eboss7' (plates 8355 and 8356).
The spectra were reduced with the current SDSS pipeline \citep{2012AJ....144..144B}.
They showed strong emission lines and weak absorption lines; see Fig. \ref{fig:all:spectra}. These clear features enabled clean redshift identification for about 75\% of the targets. 

\subsection{VIPERS data, a parent sample}
To understand the completeness properties of the ELG selection function, we use VIPERS, which has a $>90\%$ redshift determination rate for magnitudes brighter than $i<22.5$ and at redshift higher than $z>0.6$. We use its first data release, containing 57,204 slit-extracted 1d spectra and their measured and visually inspected redshifts \citep{2014A&A...562A..23G} from 45-minute-long exposures. The VIPERS data cover the fields W1 and W4 of the CFHT-LS. The eBOSS ELG pilot survey footprint and the VIPERS W1 field overlap slightly. These spectra were observed by the ESO Very Large Telescope (VLT) and VIMOS,  which is a visible (360 to 1000 nm) wide-field imager and multi-object spectrograph mounted on the Nasmyth focus B of UT3 Melipal \citep{2003SPIE.4841.1670L}. It was used in a mode with a resolution of 230 and a wavelength coverage of 550nm - 950nm.

\section{Redshift determination}
\label{sec:BOSS:ELG:redshifts}

The SDSS/BOSS pipeline \citep{2012AJ....144..144B} fits the redshifts of the observed spectra. In this section, we investigate the precision obtained on the estimation of the redshifts and the rate of catastrophic errors.
To measure the two-point correlation function in redshift space for BAO in the redshift range 0.6 to 1.1, we require a redshift precision better than 300 km s$^{-1}$ and a share of redshifts with an error larger than 1000 km s$^{-1}$ to be smaller than 1\% (the so-called catastrophic redshifts). Because eBOSS is
required to observe very many spectra, it is necessary to have reliable redshifts estimated automatically.

This pipeline has not yet been thoroughly tested in its ability to automatically determine ELG redshifts at $z\sim0.8$. This is therefore important to properly evaluate the efficiency of ELG selection functions.

\subsubsection*{Visual inspection}
To assess the fit redshifts, we inspect them visually to infirm or confirm the results. In total, we visually inspected 13,450 of the 21,500 spectra reduced by the same pipeline to assess the plausibility of the redshift assignment. We did not inspect all of them because this is a highly time-consuming task.
In this task, we profit from the previous ELG observations by BOSS that have been visually inspected (11,650 spectra), and in addition, we inspected two of the pilot survey plates (8123 and 8130, i.e.. 1,800 spectra). Each plate was assigned two inspectors. Using the same software, inspectors produced a file that contained for each spectrum the category of the object (star, galaxy, qso, unknown) and a confidence level (from 0, or low confidence, to 5, or high confidence) and the redshift. Any disagreements between inspectors were discussed between them and a truth table was constructed based on the inspection result for each plate. For confidence levels $>3$ the inspectors always agreed. Disagreements were rare and occurred when the signal-to-noise ratios in the emission lines were very low. In most of these cases, these objects were categorized as unknown to combine
inspections into the truth table. We note that when an object was classified as unknown,  $z_\mathrm{inspection}$ was set to -1.
This classification follows conventions used by spectroscopic surveys such as VVDS, DEEP2, or zCOSMOS \citep{2013A&A...559A..14L,2013ApJS..208....5N,Lilly_2009} with a slightly higher degree of detail and without comparison to photometric redshifts. 

\subsubsection*{Redshift fits from the pipeline}
The pipeline fits templates (stars, galaxies, and quasars) to the observations and outputs the most likely redshift, $Z$, an estimate of the error associated with the redshift, $Z_{ERR}$, and warning flags ZWARNING\footnote{\url{http://www.sdss.org/dr12/algorithms/bitmasks/#ZWARNING}} \citep{2012AJ....144..144B}. The best-fit template gives the object its class: star, galaxy, or QSO. The ZWARNING values of interest are 0, meaning that there was no problem during the fit, and 4, meaning that there is a small difference in $\chi^2$ between the first and the second best fit. The other values of ZWARNING enable tracking issues, but mean that we the redshift output should not be considered.

\subsubsection*{A posteriori redshift flags, zQ, and zCont}
The reliability of the redshift of an emission-line galaxy is mainly correlated with the signal-to-noise ratio (S/N) of the detection of the line(s): 
\begin{equation}
\rm S/N_{line} \; = \; fitted \; flux_{line} \; / \; error \; fitted \; flux_{line}.
\end{equation}
In some cases, it can also be correlated to the detection of a small 4000 ${\rm\AA}$ break. 

We use the redshift fit by the pipeline to classify the spectra according to the strength of their emission lines, using a flag zQ (where Q stands for quality), and to the features seen in the continuum, using a flag zCont. 
We define the flags in Table \ref{table:zFlag:EL}. In this classification, we consider a line with $3\leq S/N<5$ as a low S/N detection and a line with $S/N\geq5$ as a high S/N detection. To give an order of magnitude, the S/N value in the fitted line flux correspond roughly to the S/N value in the pixel that contains the maximum of the line.

In the following we mean the a posteriori redshift flags and not those from the inspection when we refer to redshift flags.

\begin{table}
\caption{Redshift flags. Low and high S/N correspond to $3\leq S/N<5$ and $5\leq S/N$, respectively. The \OII 3727, 3729 emission line doublet is a special case. It is sometimes observed as a blended doublet, sometimes only as a single line. The CLASS = `STAR' category means that the best fit of the pipeline is based on a stellar template.}
\begin{center}
\begin{tabular}{cc}
\hline  \hline
zQ & meaning \\ \hline
-2 & $Z\_ERR>0.005(1+Z)$ or ZWARNING$\neq$0 or 4\\
-1 & CLASS = `STAR' \\
1. & one line at low S/N \\
1.5 & two lines at low S/N \\
2. & one line at high S/N \\
2.5 & three or more lines at low S/N \\
3. & one line at high S/N and at least one line at low S/N \\
3.5 & \OII 3728 at high S/N \\
4. & two lines at high S/N \\
4.5 & three or more lines at high S/N \\
0 & none of the conditions above are met \\
\hline \hline
zCont & meaning \\
\hline
2.5 & magnitude $u$ or $g$ or $i$ or $z<19.5$ \\
2. & \hspace{1cm} " \hspace{1cm}  $<20$ \\
1.5 & \hspace{1cm} " \hspace{1cm}  $<20.5$ \\
1 & $>$3 lines with the continuum detected at S/N 10 \\
0.5 & $>$3 \hspace{1cm} " \hspace{1cm} S/N 8 \\
0 & none of the conditions above are met \\
\hline \hline
\end{tabular}
\end{center}
\label{table:zFlag:EL}
\end{table}%

\begin{table}
\begin{center}
\caption{Classification of 13,450 visually inspected redshifts per category of redshift flag. The percentages do not depend on magnitude or on selection. A redshift is counted in the N agree column when $|z_\mathrm{inspection}-z_\mathrm{pipeline}|/(1+z_\mathrm{pipeline})\leq0.005$. For zQ=-1, the fraction showed by N agree means that the inspectors also found that it was a star.
}
\label{table:summary:zInspection}
\begin{tabular}{r r rrr}
\hline \hline
\multicolumn{2}{c}{Flags} & \multicolumn{3}{c}{Inspection result} \\
zQ & zCont & N & N$_\mathrm{agree}$ & \%  \\ \hline

-2.0  &  0.0  &  573  &  71  &  12.39  \\
-2.0  &  0.5  &  13  &  4  &  30.77  \\
-2.0  &  1.0  &  68  &  12  &  17.65  \\
\hline
-1.0  &  0.0  &  62  &  6  &  9.68  \\
-1.0  &  0.5  &  7  &  2  &  28.57  \\
-1.0  &  1.0  &  397  &  193  &  48.61  \\
-1.0  &  2.5  &  29  &  28  &  96.55  \\
\hline
0.0  &  0.0  &  1453  &  490  &  33.72  \\
0.0  &  0.5  &  43  &  23  &  53.49  \\
0.0  &  1.0  &  115  &  70  &  60.87  \\
0.0  &  1.5  &  76  &  70  &  92.11  \\
0.0  &  2.0  &  22  &  18  &  81.82  \\
0.0  &  2.5  &  12  &  12  &  100.00  \\
\hline
1.0  &  0.0  &  618  &  337  &  54.53  \\
1.0  &  0.5  &  62  &  55  &  88.71  \\
1.0  &  1.0  &  177  &  160  &  90.40  \\
1.0  &  1.5  &  61  &  60  &  98.36  \\
1.0  &  2.0  &  27  &  26  &  96.30  \\
1.0  &  2.5  &  16  &  16  &  100.00  \\
\hline
1.5  &  0.0  &  274  &  187  &  68.25  \\
1.5  &  0.5  &  64  &  53  &  82.81  \\
1.5  &  1.0  &  146  &  137  &  93.84  \\
1.5  &  1.5  &  36  &  36  &  100.00  \\
1.5  &  2.0  &  18  &  18  &  100.00 \\
1.5  &  2.5  &  7  &  7  &  100.00 \\
\hline
2.0  &  0.0  &  218  &  175  &  80.28  \\
2.0  &  0.5  &  6  &  2  &  33.33  \\
2.0  &  1.0  &  10  &  7  &  70.00  \\
2.0  &  1.5  &  26  &  26  &  100.00 \\
2.0  &  2.0  &  12  &  12  &  100.00 \\
2.0  &  2.5  &  3  &  2  &  66.67  \\
\hline
2.5  &  0.0  &  136  &  118  &  86.76  \\
2.5  &  0.5  &  31  &  30  &  96.77  \\
2.5  &  1.0  &  64  &  58  &  90.62  \\
2.5  &  1.5  &  27  &  27  &  100.00 \\
2.5  &  2.0  &  16  &  16  &  100.00 \\
2.5  &  2.5  &  2  &  2  &  100.00 \\
\hline
3.0  &  0.0  &  205  &  175  &  85.37  \\
3.0  &  0.5  &  22  &  20  &  90.91  \\
3.0  &  1.0  &  63  &  61  &  96.83  \\
3.0  &  1.5  &  46  &  46  &  100.00 \\
3.0  &  2.0  &  24  &  23  &  95.83  \\
3.0  &  2.5  &  10  &  9  &  90.0  \\
\hline
3.5  &  0.0  &  1567  &  1516  &  96.75  \\
3.5  &  0.5  &  335  &  332  &  99.10  \\
3.5  &  1.0  &  612  &  612  &  100.00 \\
3.5  &  1.5  &  272  &  271  &  99.63  \\
3.5  &  2.0  &  126  &  126  &  100.00 \\
3.5  &  2.5  &  28  &  28  &  100.00 \\
\hline
4.0  &  0.0  &  721  &  721  &  100.00 \\
4.0  &  0.5  &  251  &  251  &  100.00 \\
4.0  &  1.0  &  560  &  560  &  100.00 \\
4.0  &  1.5  &  264  &  264  &  100.00 \\
4.0  &  2.0  &  99  &  99  &  100.00 \\
4.0  &  2.5  &  29  &  29  &  100.00 \\
\hline
4.5  &  0.0  &  1036  &  1036  &  100.00 \\
4.5  &  0.5  &  406  &  406  &  100.00 \\
4.5  &  1.0  &  1263  &  1263  &  100.00 \\
4.5  &  1.5  &  398  &  398  &  100.00 \\
4.5  &  2.0  &  132  &  132  &  100.00 \\
4.5  &  2.5  &  54  &  54  &  100.00 \\ \hline
\end{tabular}
\end{center}
\end{table}

\subsection{Redshift reliability}
In this section we compare the inspections, the outputs of the pipeline, and the flags zQ and zCont. We determine a scheme that makes more inspections redundant and provides reliable redshifts to the collaboration.

To gather the three sources of information into a single table, we convert the results from the inspections into a single decision: the inspector agrees or disagrees with the redshift proposed by the pipeline: $|z_\mathrm{inspection}-z_\mathrm{pipeline}|/(1+z_\mathrm{pipeline})\leq0.005$. 
Table \ref{table:summary:zInspection} presents the results. The higher the flag value, the higher the agreement rate, thus the flags are a good estimator of the redshift quality.
Based on Table \ref{table:summary:zInspection}, we can create a criterion to select the largest number of redshifts for the
highest tolerated error rate, that is, we investigate the trade-off between the total amount of redshift considered and the fraction of redshifts that are catastrophic. 

For the purpose of eBOSS clustering analysis, we select galaxies with a pipeline redshift in the range $0.7<z<1.1$ that are brighter than $g<22.8$ (using DES or CFHT photometry, in total, we have 2660 such galaxies to study this trade-off).

We find that the following criterion fulfills the eBOSS/ELG redshift quality requirement:
\begin{equation}
\rm zQ\geq2\;  or \; (zQ\geq1\;  and\;  zCont>0)\;  or \; (zQ\geq0\;  and \;  zCont\geq2.5). 
\label{zOK:selection}
\end{equation}

The criterion excludes 278 objects and keeps 2382 objects. Of the 2382, the inspection disagrees with 22 redshifts, which corresponds to a 1\% share. Of the 278 excluded objects, the inspection agrees with 147 redshifts, which is a 52\% share. Future pipeline improvement could therefore lead to an improvement of 5.5\% in efficiency
at most. For a clustering analysis, we were unable to consider all the redshifts provided by the pipeline because it produces a fraction of incorrect redshifts of about 6\%. In other words, we need to discard redshifts of lower quality to obtain a purer sample. With the current observations, we cannot determine the effect of this poor-redshift exclusion on the redshift distribution and on the clustering measurement.

\subsubsection{Comparison with VIPERS redshifts}

In the eBOSS ELG test plates, 383 match redshifts in the VIPERS DR1 field W1. We find 370 (97.6\%) redshifts in agreement (with $dz/(1+z)\leq0.005$) and 13 redshifts in disagreement (with $dz/(1+z)>0.005$). Ten out of the 13 galaxies have a low-quality flag in VIPERS and a high-quality flag in eBOSS. A second visual inspection of these ten eBOSS redshifts indicates that they are correct. Three out of 13 have a low-quality flag in both VIPERS and eBOSS: both redshifts may not be reliable. The last category is $3/383=0.78\%$ of the total.

\subsubsection{Redshift efficiency and fiber number}
As shown in Fig. 9 of \citet{2012AJ....144..144B}, the redshift efficiency decreases for the fiber numbers around 0, 500, and 1000 because they are the most off-center with respect to the spectrograph camera optics. The ELGs demonstrate the same trends.

\subsection{Line confusion}
If the continuum is not detected (zCont$=0$), then the redshifts
in classes zQ= 1, 2, and 3.5 only rely on the detection of a single emission line (SEL hereafter).
The BOSS spectrograph covers 3,600\AA\; to 10,400\AA,  and prominent lines with similar strengths are 
\begin{itemize}
\item \Ha ($\lambda$6564) detectable at $\rm {\it z<z}^{max}_{H\alpha}=0.584$,
\item \OIII ($\lambda$5007) detectable at $\rm {\it z<z}^{max}_{[{\sc O_{III}}]}=1.077$, 
\item \Hb ($\lambda$4862) detectable at $\rm {\it z<z}^{max}_{[{\sc H\beta}]}=1.139$,
\item \OII ($\lambda\lambda$3727,3729) detectable at $\rm {\it z<z}^{max}_{[{\sc O_{II}}]}=1.790$.
\end{itemize}
At redshift 1.8, the Lyman $\alpha$ line is at $\lambda=3,400 \AA$ and cannot be detected. The eBOSS survey will first observe quasar targets with a similar limiting magnitude as ELGs. The quasar sampling is quite complete in particular for bright Lyman $\alpha$ quasars \citep{2015ApJS..221...27M,2016A&A...587A..41P}. The remaining quasar contamination of the ELG sample will be very small ($<0.5\%$) and cannot be quantified with current data. We therefore examine the following possible confusions:
\begin{equation}
\rm \lambda^{detection}_{line} = (1+z_{H\alpha})\lambda_{H\alpha} =(1+z_{H\beta})\lambda_{H\beta}  =(1+z_{[{\sc O_{III}}]})\lambda_{[{\sc O_{III}}]} = (1+z_{[{\sc O_{II}}]}) \lambda_{[{\sc O_{II}}]}
.\end{equation}
We set aside the case zQ=3.5 and zCont=0 because the \OII doublet is sometimes seen as a blended doublet and therefore provides more information than a single line detection.

The set of observed SEL redshifts as a function of the emission line and S/N is detailed in Table \ref{table:SEL:redshifts}. At S/N$\geq5$, all SEL redshifts are primarily based on prominent lines (6 exceptions out of 3,329), whereas for the $3\leq S/N<5$ SEL redshifts, we encounter a variety of line detections (284 exceptions out of 855). 

We re-inspected the low and high S/N line detections that are not in the set of prominent lines. We found that these lines are fit on residuals of the sky subtraction.
The low and high S/N line detections found to be prominent lines are not convincing either, and line confusion is possible. 
For this reason, we exclude the classes (zQ=1 or 2) and zCont=0 from the pool of reliable redshifts in the reliable redshift selection criterion.

\begin{table}
\caption{Distribution of the SEL redshifts in the complete BOSS/eBOSS ELG sample and in the fiducial eBOSS target selections decam 180, see Sect. \ref{sec:decamTS} for the definition.}
\begin{center}
\begin{tabular}{l rrr }
\hline \hline
line & N & decam 180 &  \\ 
 \multicolumn{3}{c}{855 ELG, $3\leq S/N<5$ (zQ=1 and zCont=0)} \\ \hline
[Ar$_\textsc{III}$] ($\lambda$ 7137 ) &  8  &  0   \\ 
H$\epsilon$ ($\lambda$ 3970 ) &  33  &  0   \\ 
H$\delta$ ($\lambda$ 4102 ) &  17  &  0    \\ 
H$\gamma$ ($\lambda$ 4341 ) &  34  &  2   \\ 
H$\beta$ ($\lambda$ 4862 ) &  12  &  0  \\ 
H$\alpha$ ($\lambda$ 6564 ) &  39  &  0   \\ 
He$_\textsc{II}$ ($\lambda$ 4686 ) &  14  &  1    \\ 
He$_\textsc{II}$ ($\lambda$ 5411 ) &  8  &  0    \\ 

[N$_\textsc{II}$] ($\lambda$ 6549 ) &  6  &  0  \\ 

[N$_\textsc{II}$] ($\lambda$ 6585 ) &  19  &  0    \\ 

[Ne$_\textsc{III}$] ($\lambda$ 3869 ) &  47  &  0    \\ 

[O$_\textsc{II}$] ($\lambda$ 6302 ) &  17  &  1    \\ 

[O$_\textsc{II}$] ($\lambda$ 6365 ) &  6  &  0    \\ 

[O$_\textsc{II}$] ($\lambda$ 3728 ) &  488  &  16    \\ 

[O$_\textsc{III}$] ($\lambda$ 4363 ) &  27  &  0   \\ 

[O$_\textsc{III}$] ($\lambda$ 4960 ) &  17  &  0   \\ 

[O$_\textsc{III}$] ($\lambda$ 5007 ) &  44  &  1  \\ 

[S$_\textsc{II}$] ($\lambda$ 6718 ) &  10  &  1   \\ 

[S$_\textsc{II}$] ($\lambda$ 6732 ) &  9  &  0    \\ 

\hline \hline
line & N & decam 180 &  \\ 
 \multicolumn{3}{c}{3329 ELG, $5\leq$S/N ((zQ=2 or 3.5) and zCont=0)}\\
\hline
H$\gamma$ ($\lambda$ 4341 ) &  1  &  0    \\ 

H$\beta$ ($\lambda$ 4862 ) &  1  &  0   \\ 

H$\alpha$ ($\lambda$ 6564 ) &  85  &  0    \\ 

[N$_\textsc{II}$] ($\lambda$ 6585 ) &  2  &  0   \\ 

[Ne$_\textsc{III}$] ($\lambda$ 3869 ) &  2  &  0   \\ 

[O$_\textsc{II}$] ($\lambda$ 3728) &  3225  &  268    \\ 

[O$_\textsc{III}$] ($\lambda$ 5007 ) &  13  &  1    \\ 

\hline
\end{tabular}
\end{center}
\label{table:SEL:redshifts}
\end{table}%

\begin{figure*}
\begin{center}
\includegraphics[type=pdf,ext=.pdf,read=.pdf,height=60mm]{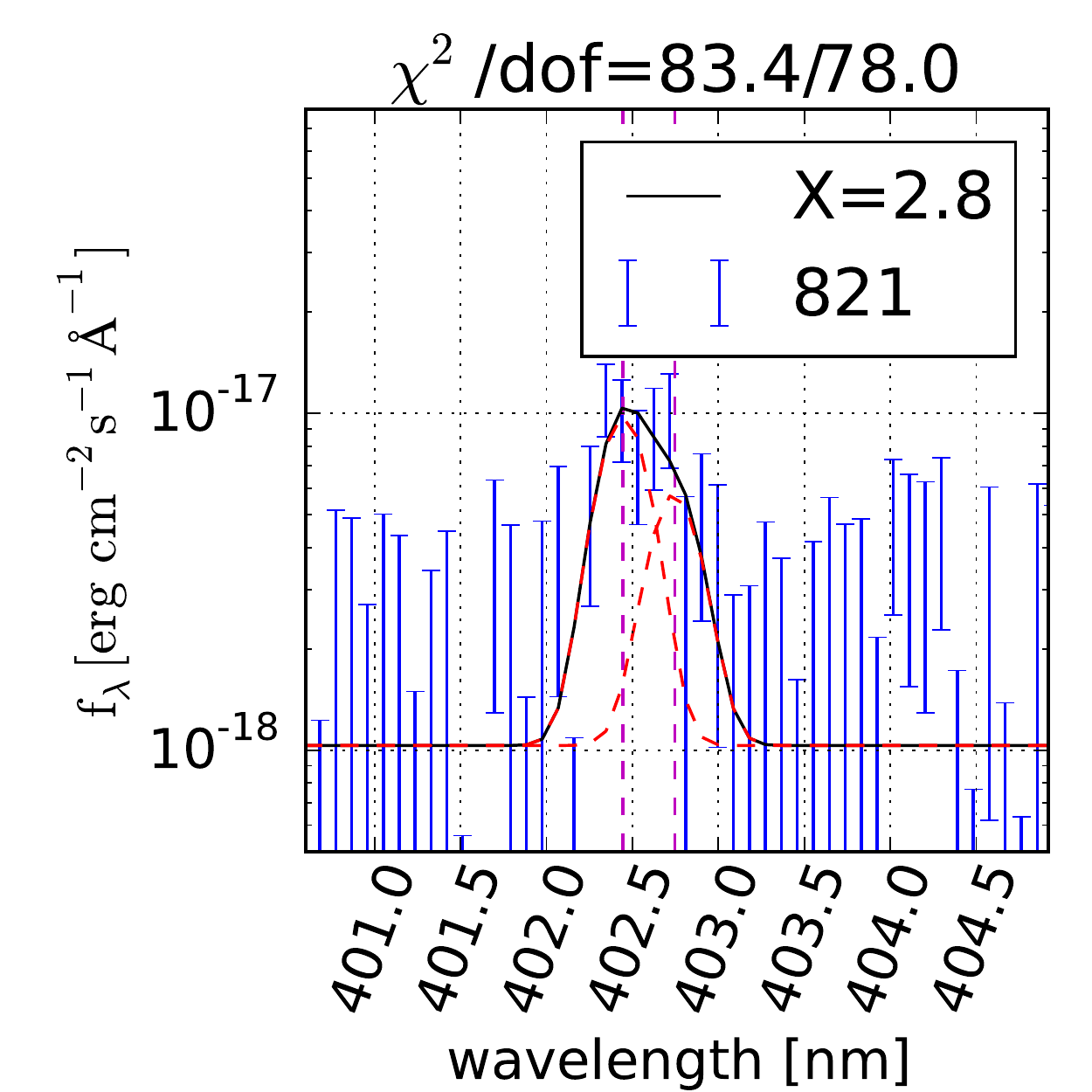}
\includegraphics[type=pdf,ext=.pdf,read=.pdf,height=60mm]{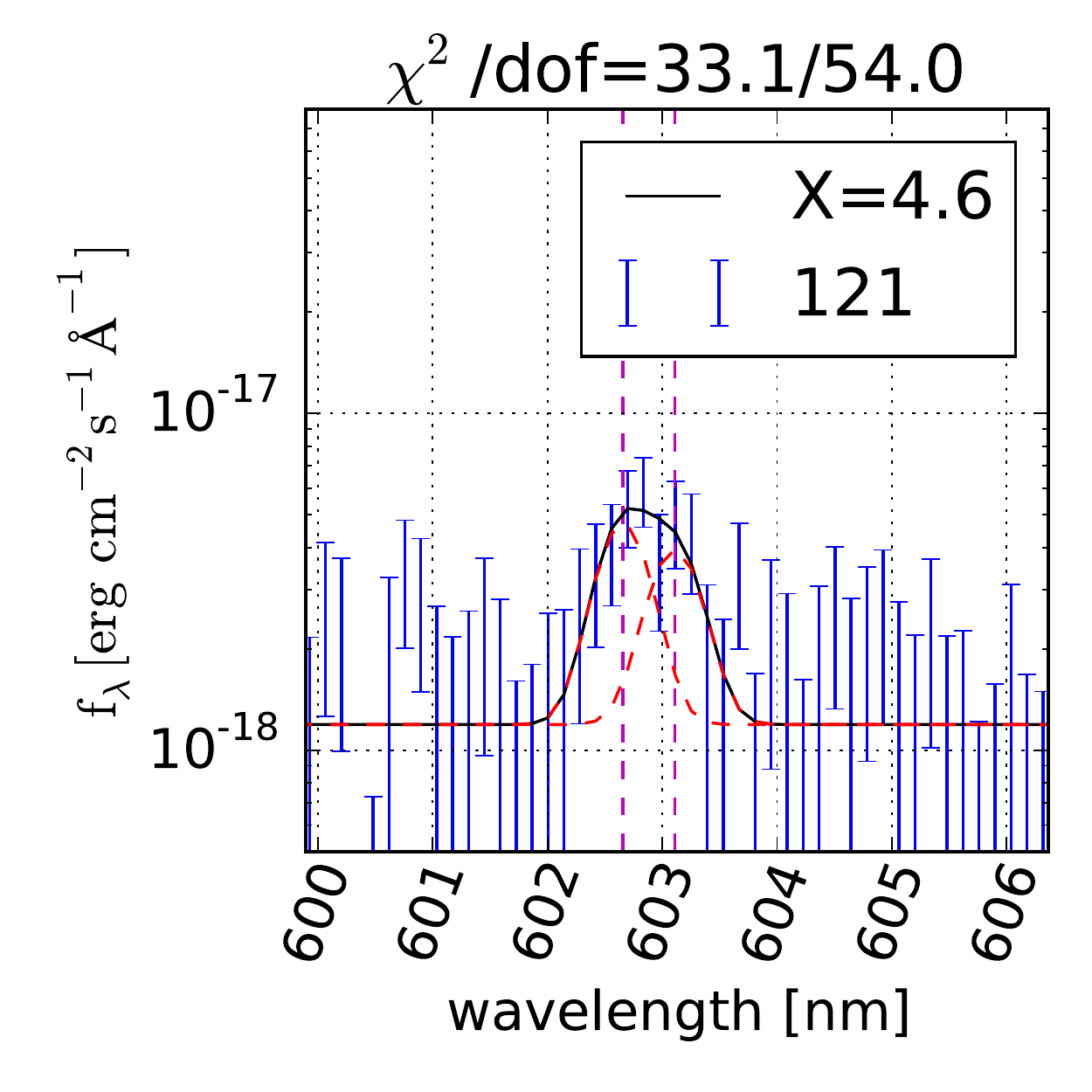}
\includegraphics[type=pdf,ext=.pdf,read=.pdf,height=60mm]{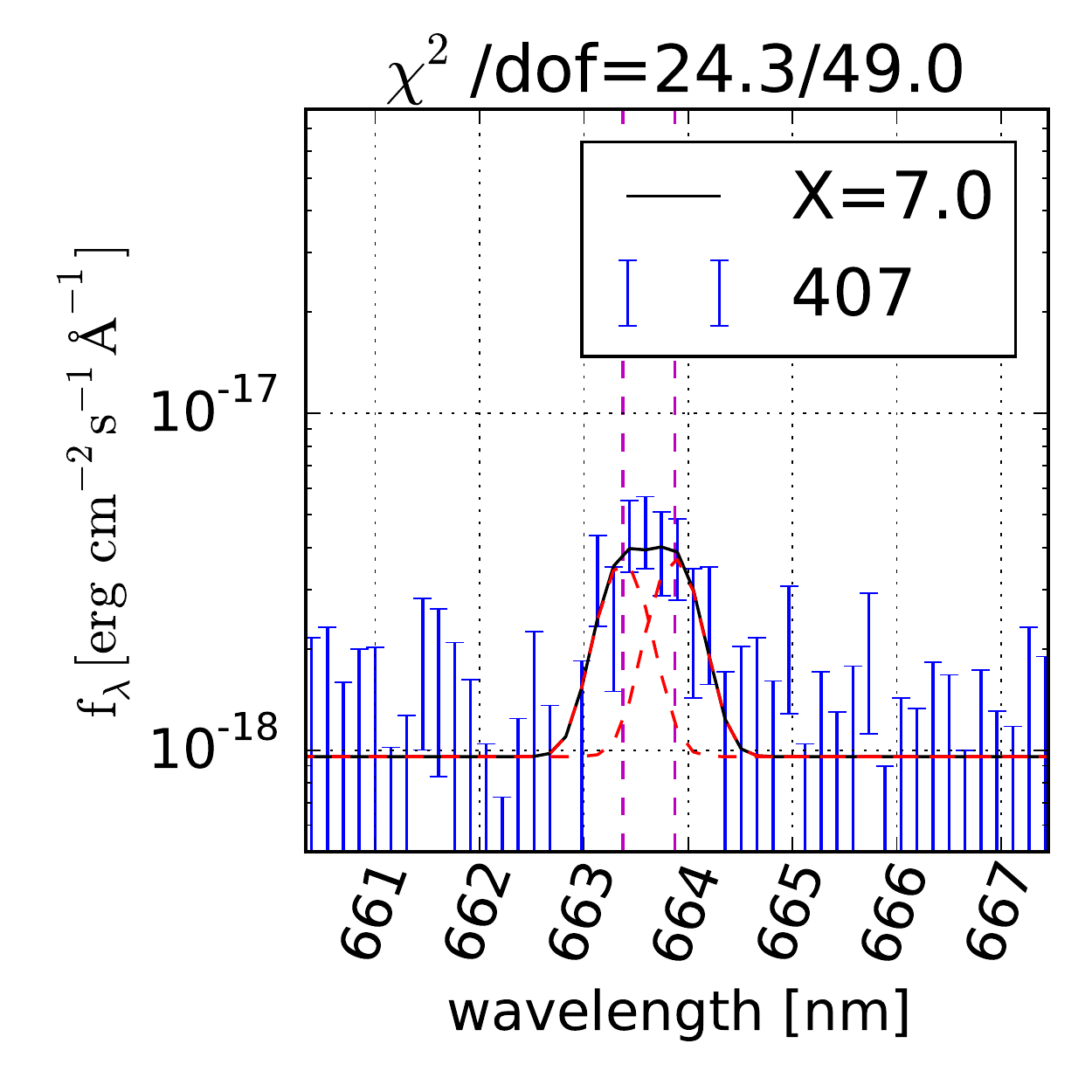}
\includegraphics[type=pdf,ext=.pdf,read=.pdf,height=60mm]{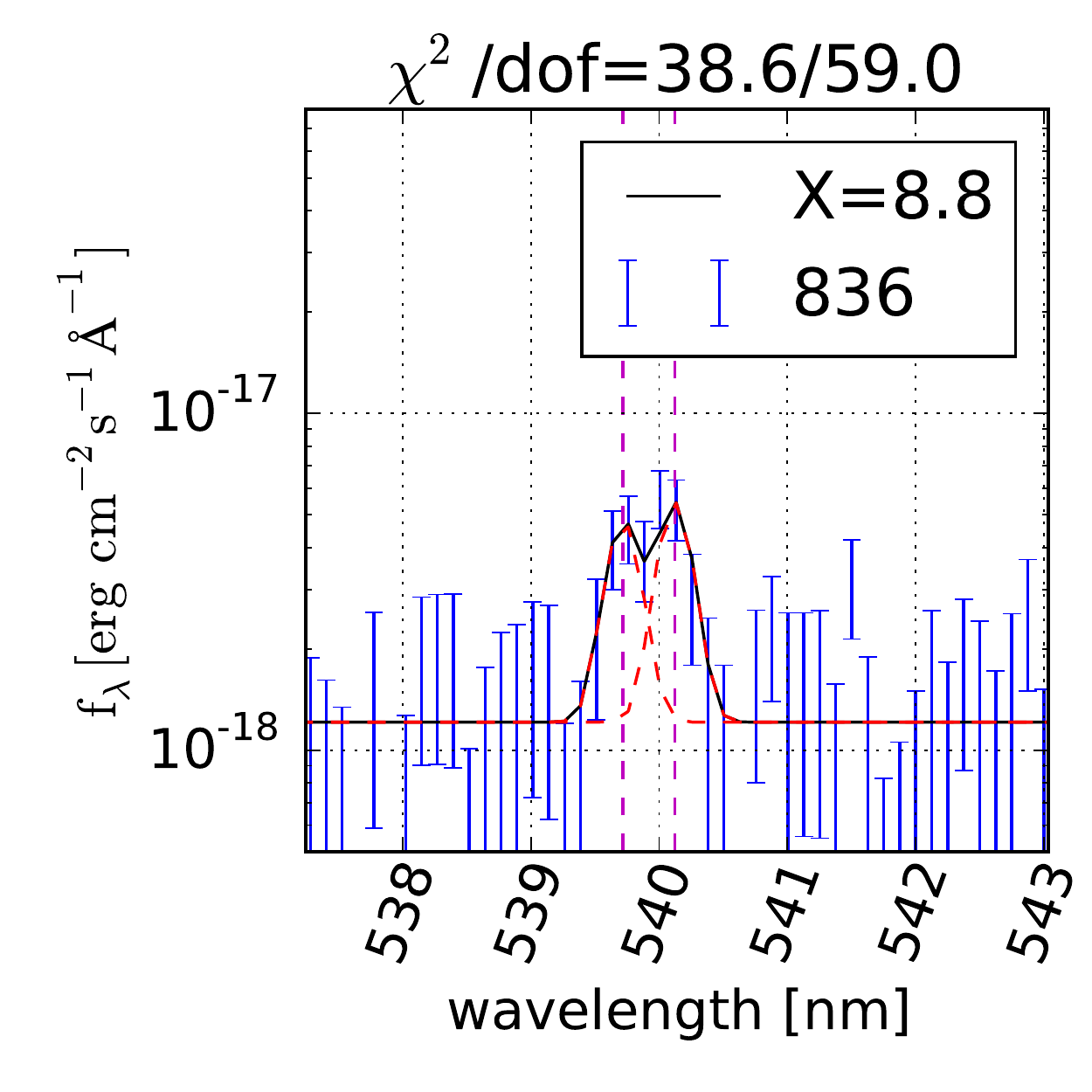}
\includegraphics[type=pdf,ext=.pdf,read=.pdf,height=60mm]{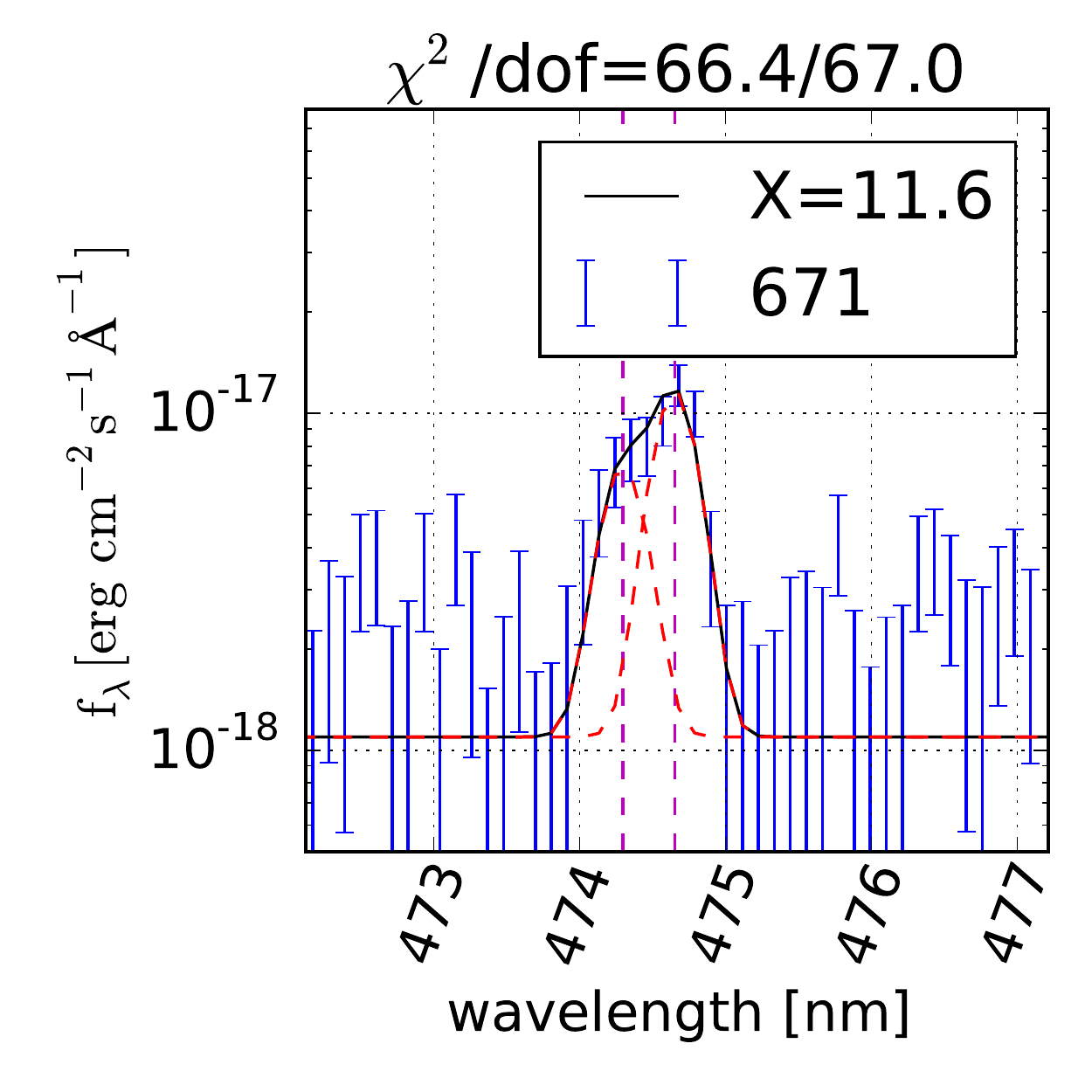}
\includegraphics[type=pdf,ext=.pdf,read=.pdf,height=60mm]{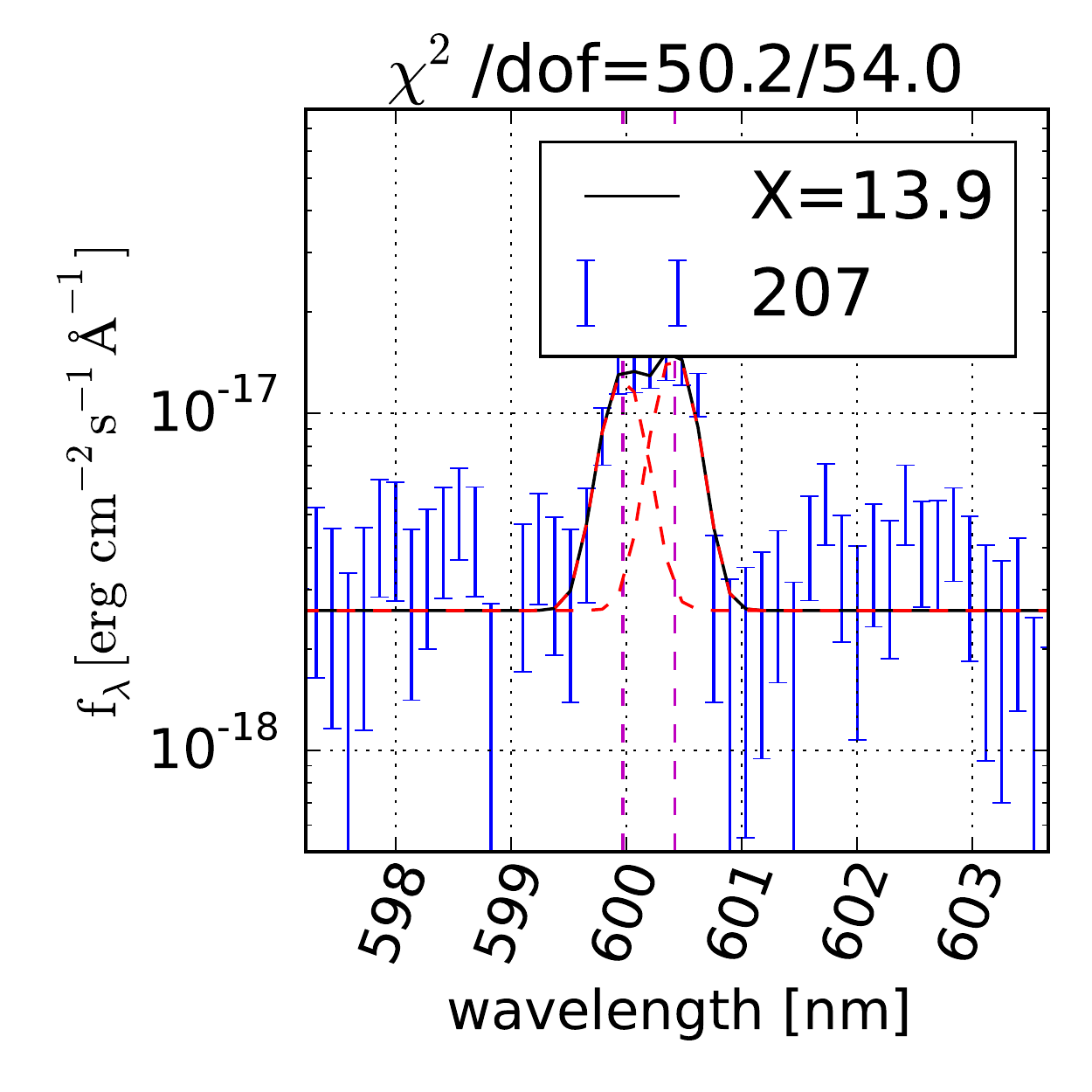}
\includegraphics[type=pdf,ext=.pdf,read=.pdf,height=60mm]{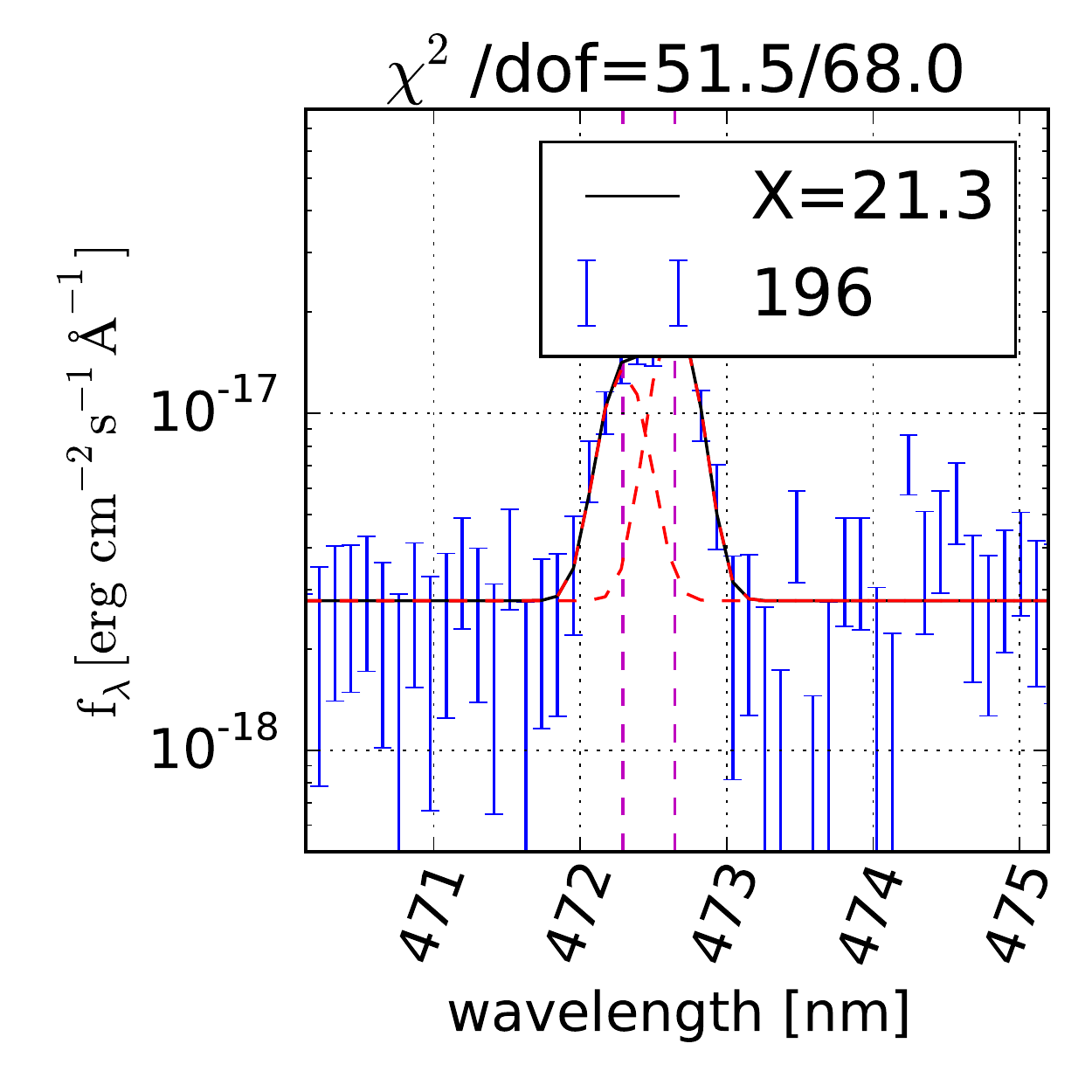}
\includegraphics[height=60mm]{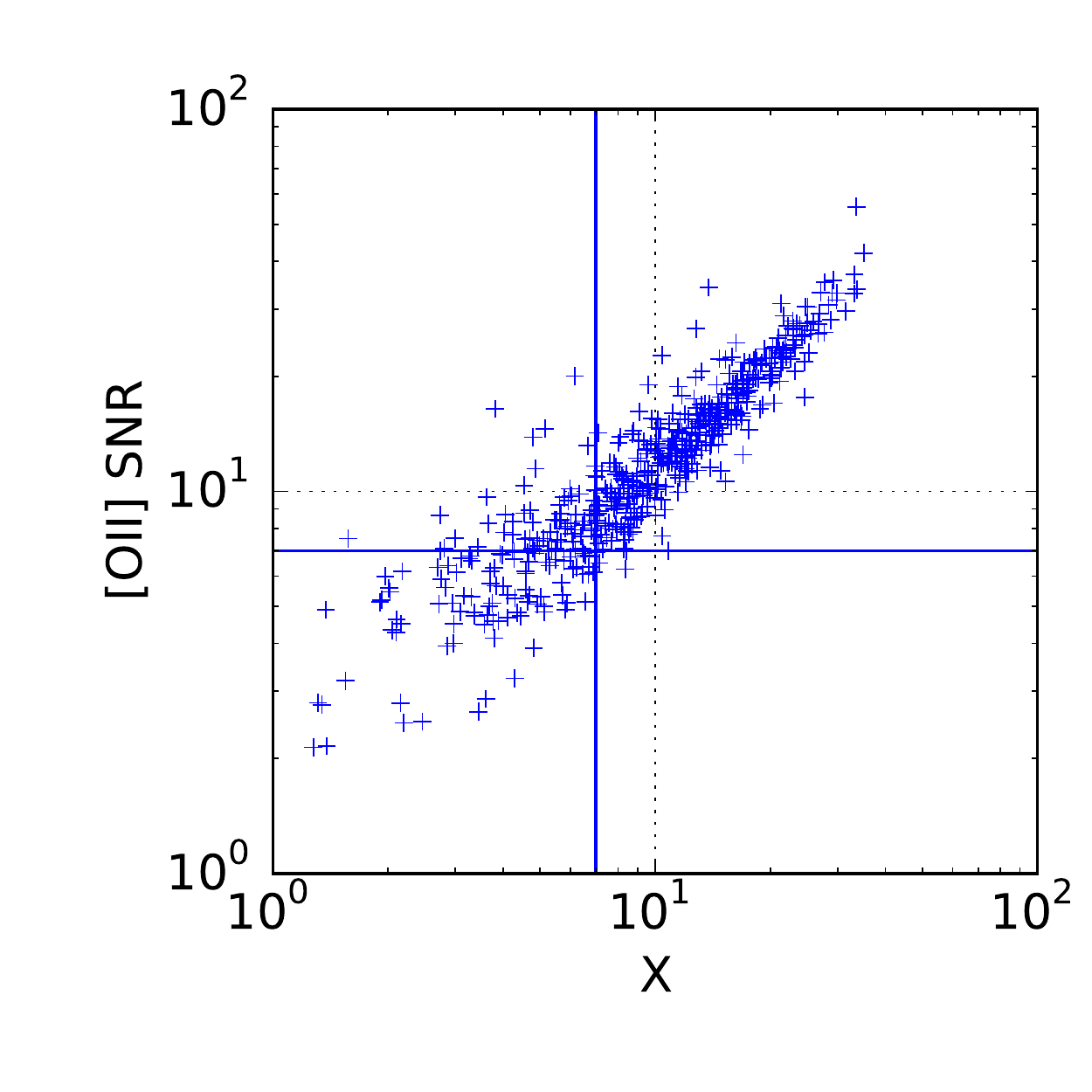}
\caption{Example of fits of the \OII doublets by a double Gaussian on some spectra from plate 8130. Each spectrum (blue error bars) is labeled by its fiber number and the fits (black solid) are characterized by the $\chi^2/$dof given above each panel. We show the two individual Gaussians as red dashes. The vertical magenta dashed lines are at $(1+z) \lambda_{\rm[OII]}$. The panels are ordered by increasing value of X. Around a value of X=7, the blended doublet starts to be a better model than a single Gaussian. The last panel shows the \OII S/N vs. X with horizontal and vertical solid lines at a value of 7. X is the S/N with which the two components of the doublets are detected.}
\label{fig:oii:doublet}
\end{center}
\end{figure*}

There are 3,225 S/N5 \OII detection cases with zQ=3.5 and zCont=0. For the \OII line, we fit 
the share of the flux in each component of the line:
\begin{equation}
\rm \alpha = flux_{\lambda 3729} / ( flux_{\lambda 3727}+flux_{\lambda 3729} ).
\end{equation}
We define $X=\alpha/\alpha_{err}$ and show in Fig \ref{fig:oii:doublet} a few \OII lines measurements and fits that span the range of X values. We find that X is correlated to the S/N detection of the line.
On average, if the doublet is detected at S/N 7, then the double Gaussian model is significantly more accurate than a simple Gaussian model.
We note that at redshift 0.8, the typical velocity dispersion in the ELG is around 70 km s$^{-1}$, so that the line width of each component in the fits is dominated by the instrumental resolution \citep{2013A&A...559A..18C}.

\subsection{Pipeline redshift error}
The previous subsection demonstrated that the redshift failures are correlated with the S/N in the lines. We now quantify this statement using the pipeline redshift errors. 

\begin{figure*}
\begin{center}
\includegraphics[height=45mm]{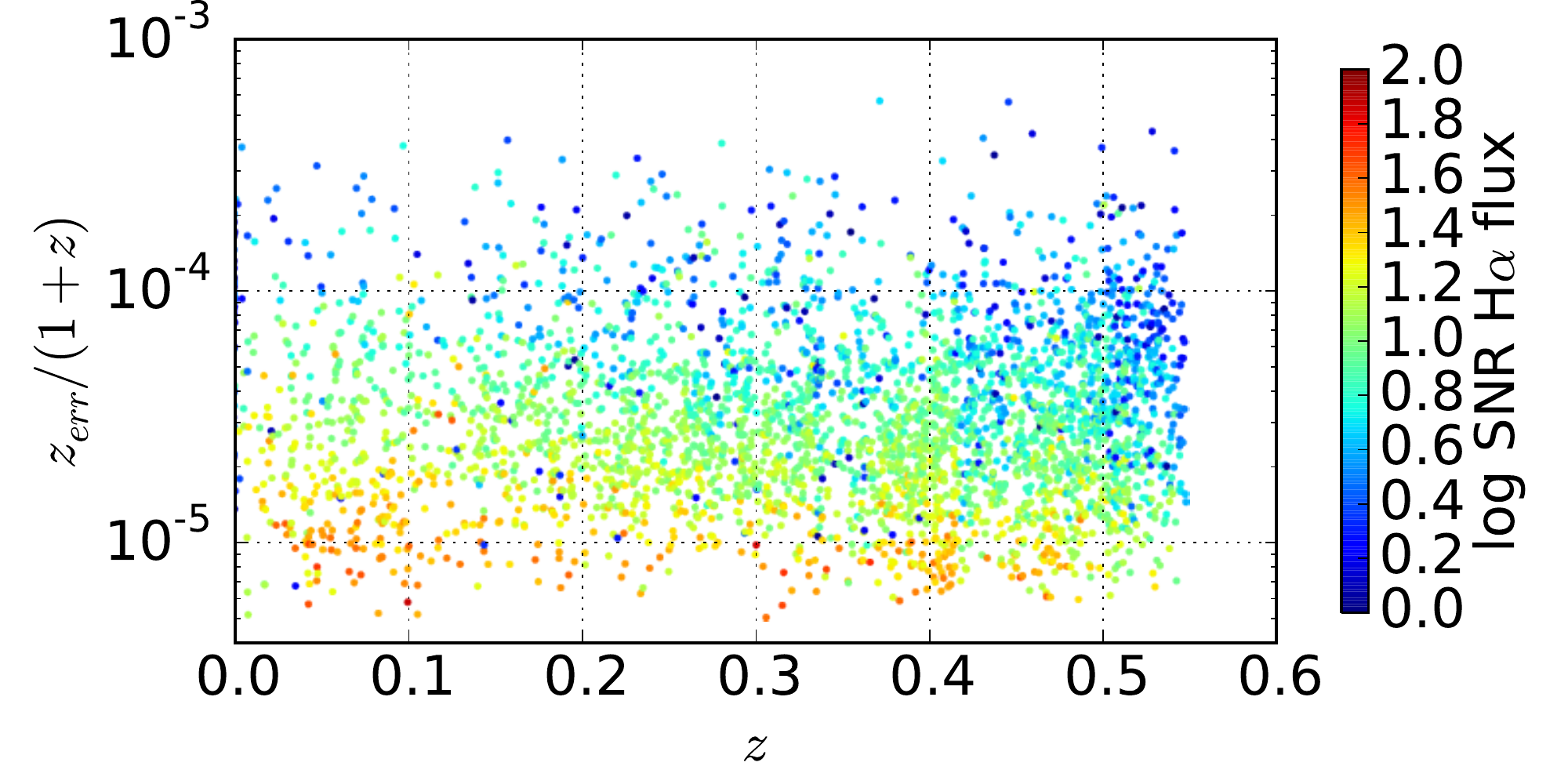}
\includegraphics[height=45mm]{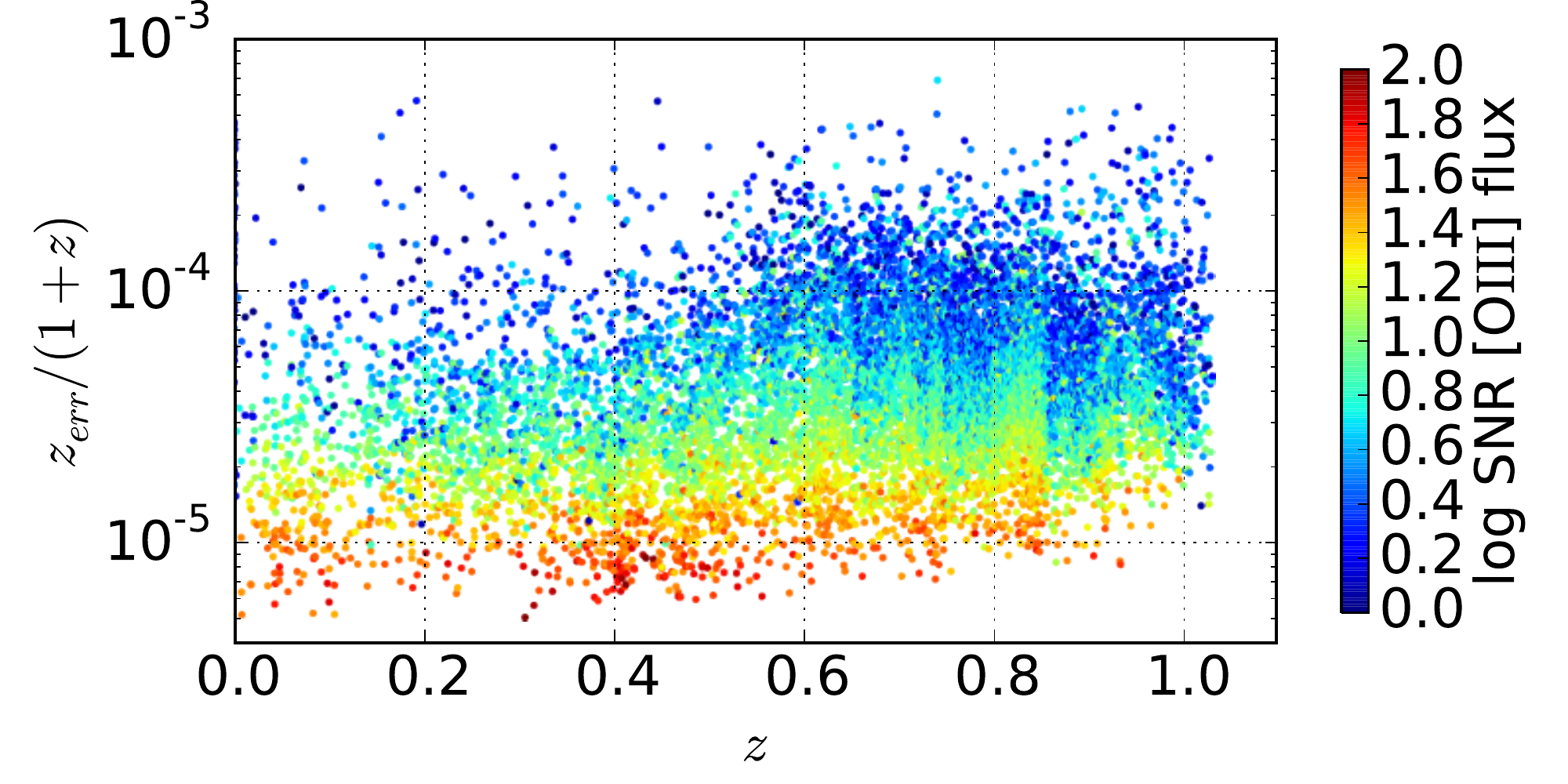}
\includegraphics[height=45mm]{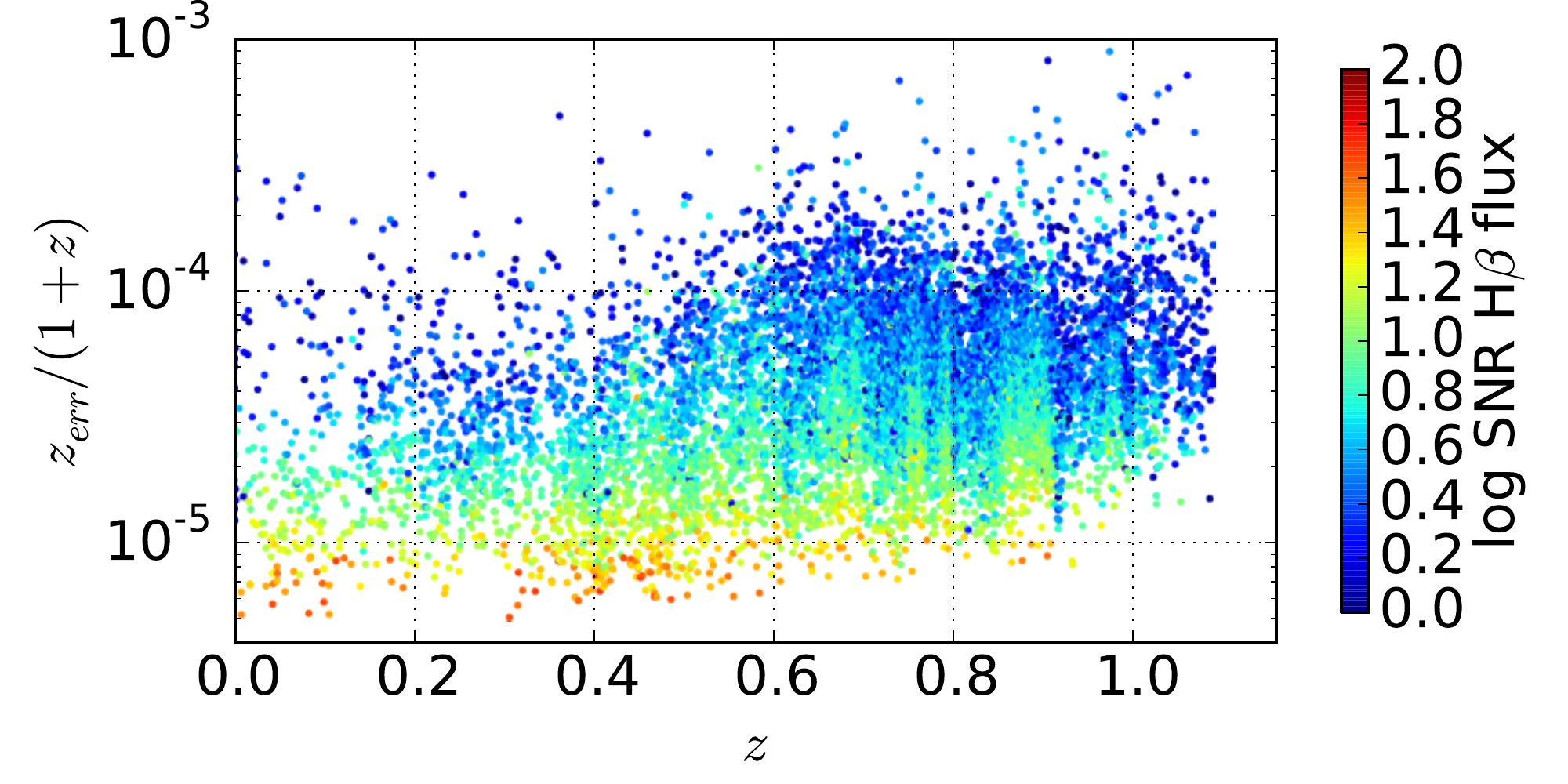}
\includegraphics[height=45mm]{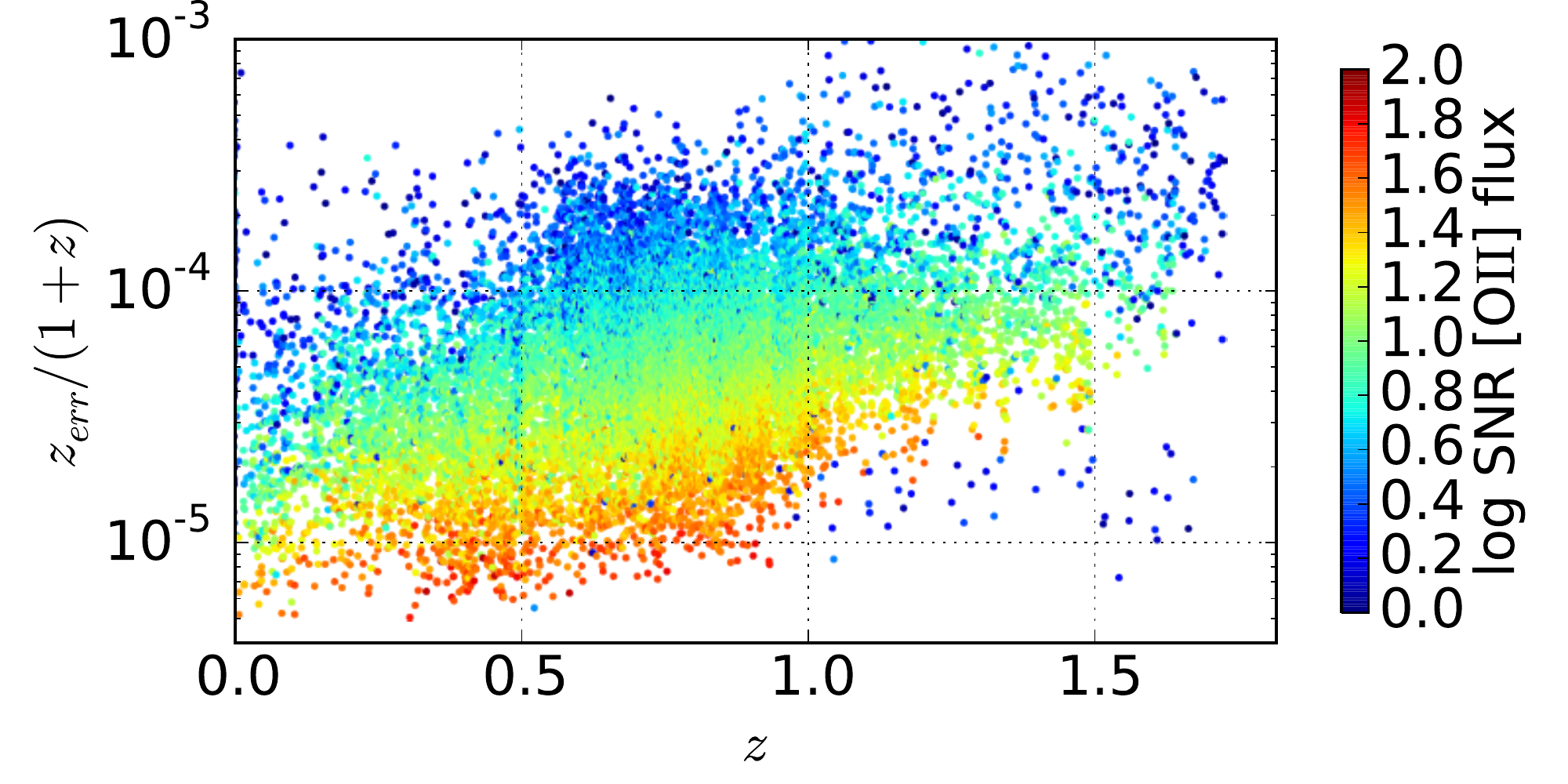} \\
\includegraphics[height=5cm]{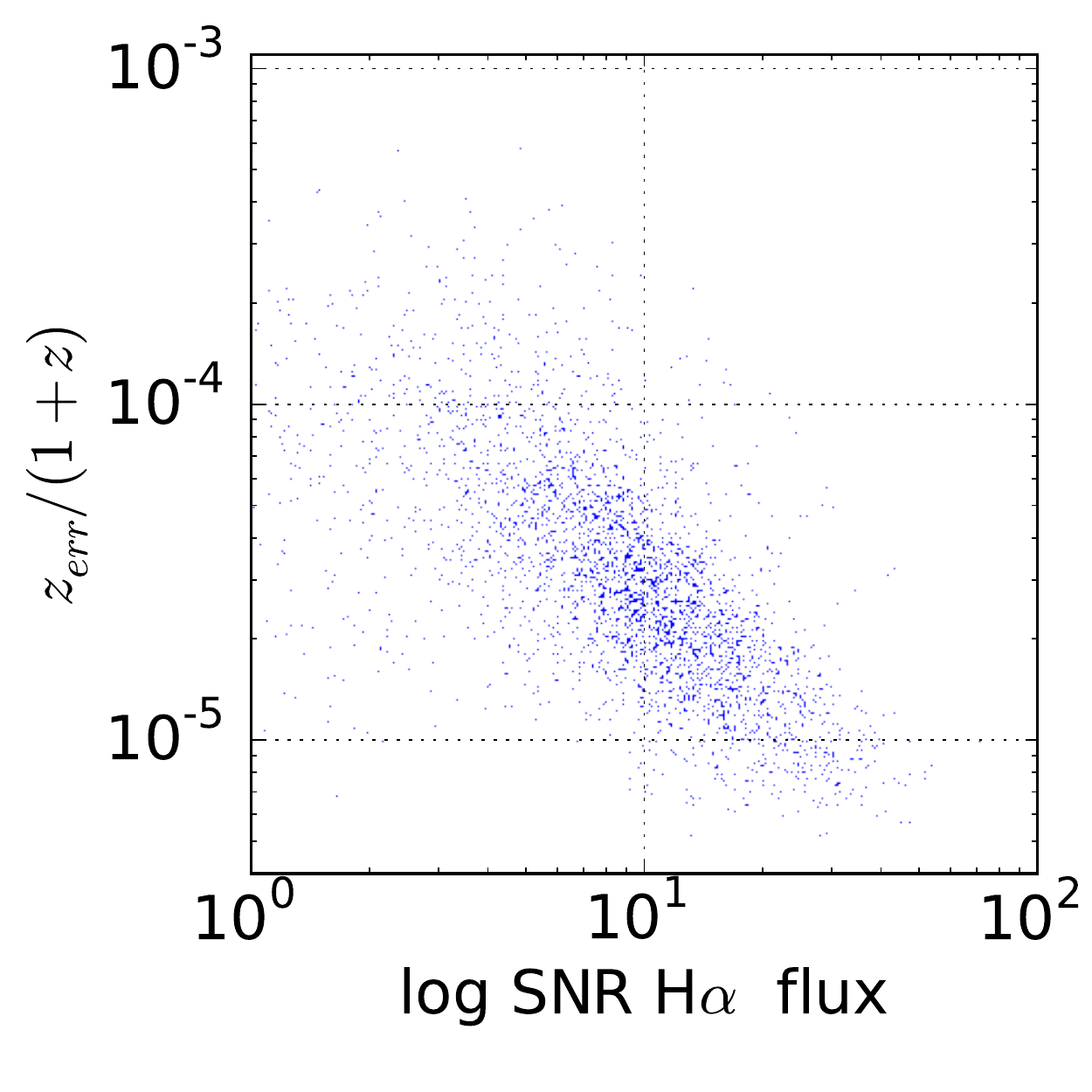}
\includegraphics[height=5cm]{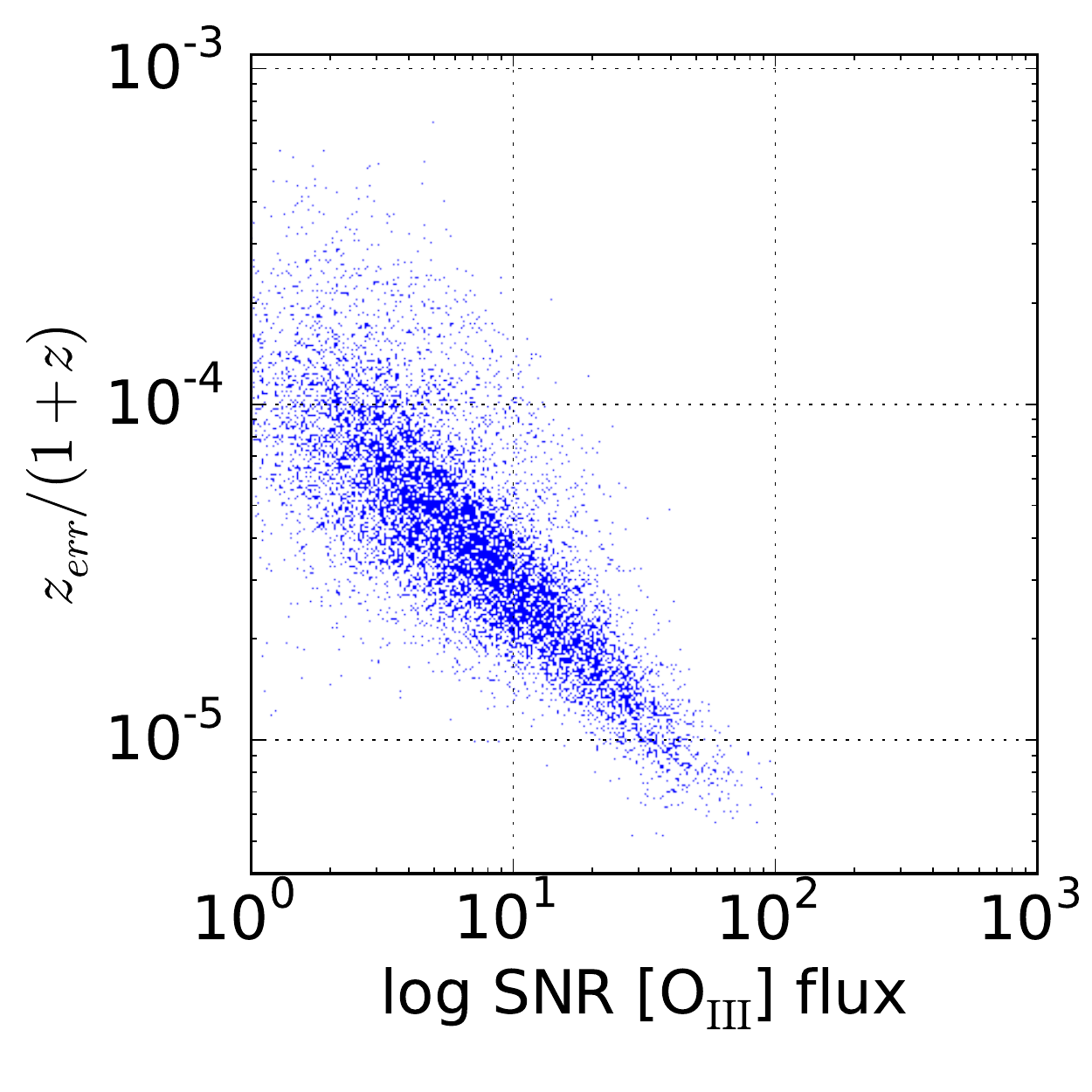}
\includegraphics[height=5cm]{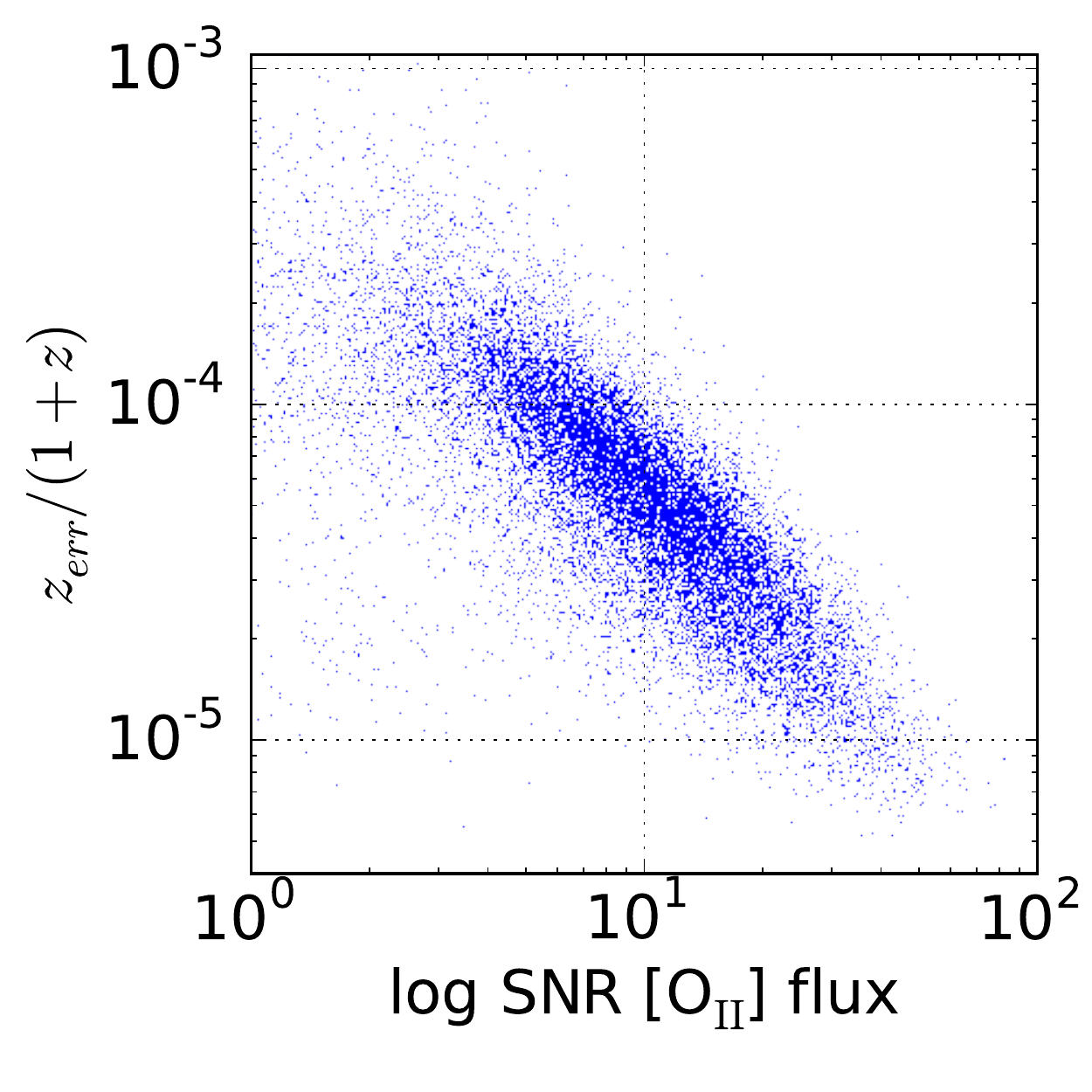}
\caption{Redshift uncertainty vs redshift for the ELGs for the \Ha, \OIII, \Hb, and \OII emission lines detected coded as a function of the log of the detection S/N. The sky emission line is imprinted on the variation in S/N with redshift. The last row presents the correlation between redshift error and line S/N detection. The redshift error is anticorrelated with the line S/N.}
\label{fig:redshift:errors:boss}
\end{center}
\end{figure*}

We consider all the redshifts (with zQ$\geq-1$) and the correlation between the detection S/N of the line, the redshift, and redshift error.
Figure \ref{fig:redshift:errors:boss} shows the redshift error as a function of redshift coded with the emission line S/N for the lines \Ha \Hb, \OII, and \OIII 5007. Regardless of redshift, the S/N of the line S/N is highly anticorrelated with the redshift error: the higher the S/N, the lower the error. The effect of the sky brightness on the redshift error is also evident: for a fixed S/N in a line, the redshift error increases and decreases as the sky brightness.
A S/N=3 in the \OII line corresponds to an average error of $z_{err} \sim (1+z)\times 10^{-4}$, suggesting that the pipeline redshift errors are within the requirements. 
To conclude, the stronger the emission lines, the more accurate the centroid of the redshift.

\section{ELG selection algorithms}
\label{sec:ELGTS:algorithms}

\begin{figure*}
\begin{center}
\includegraphics[height=0.4\textwidth]{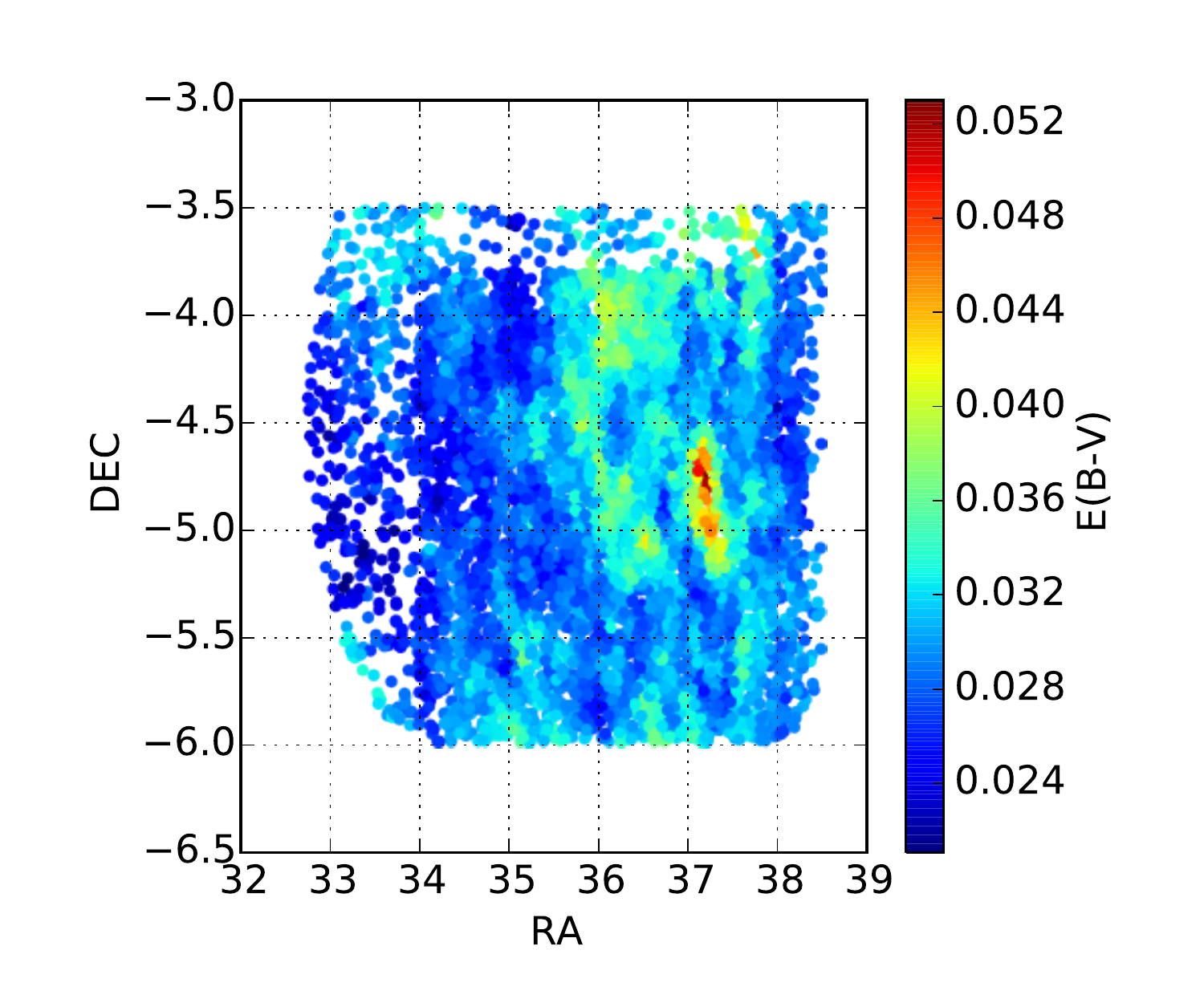}
\includegraphics[height=0.4\textwidth]{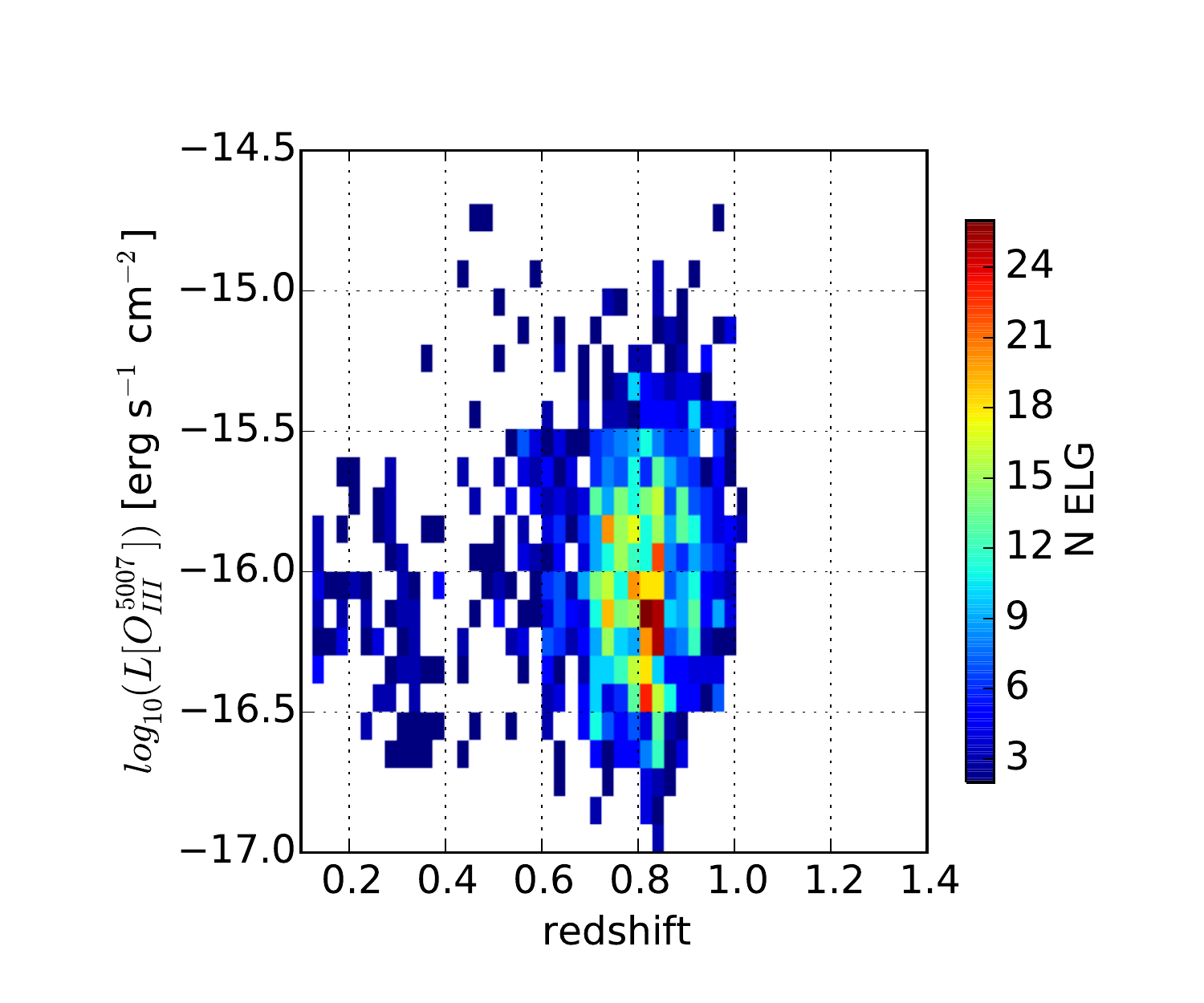} \\
\includegraphics[height=0.4\textwidth]{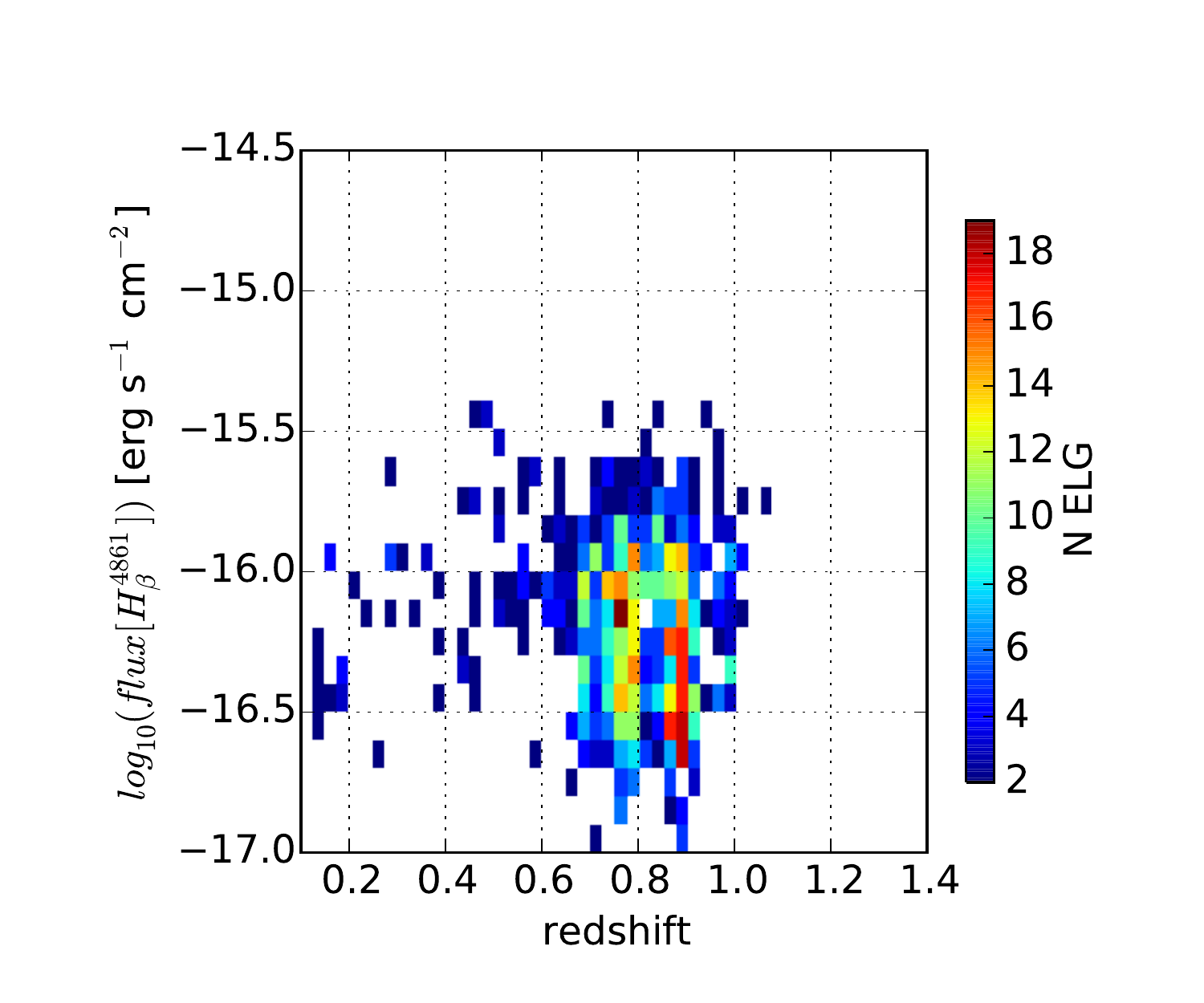}
\includegraphics[height=0.4\textwidth]{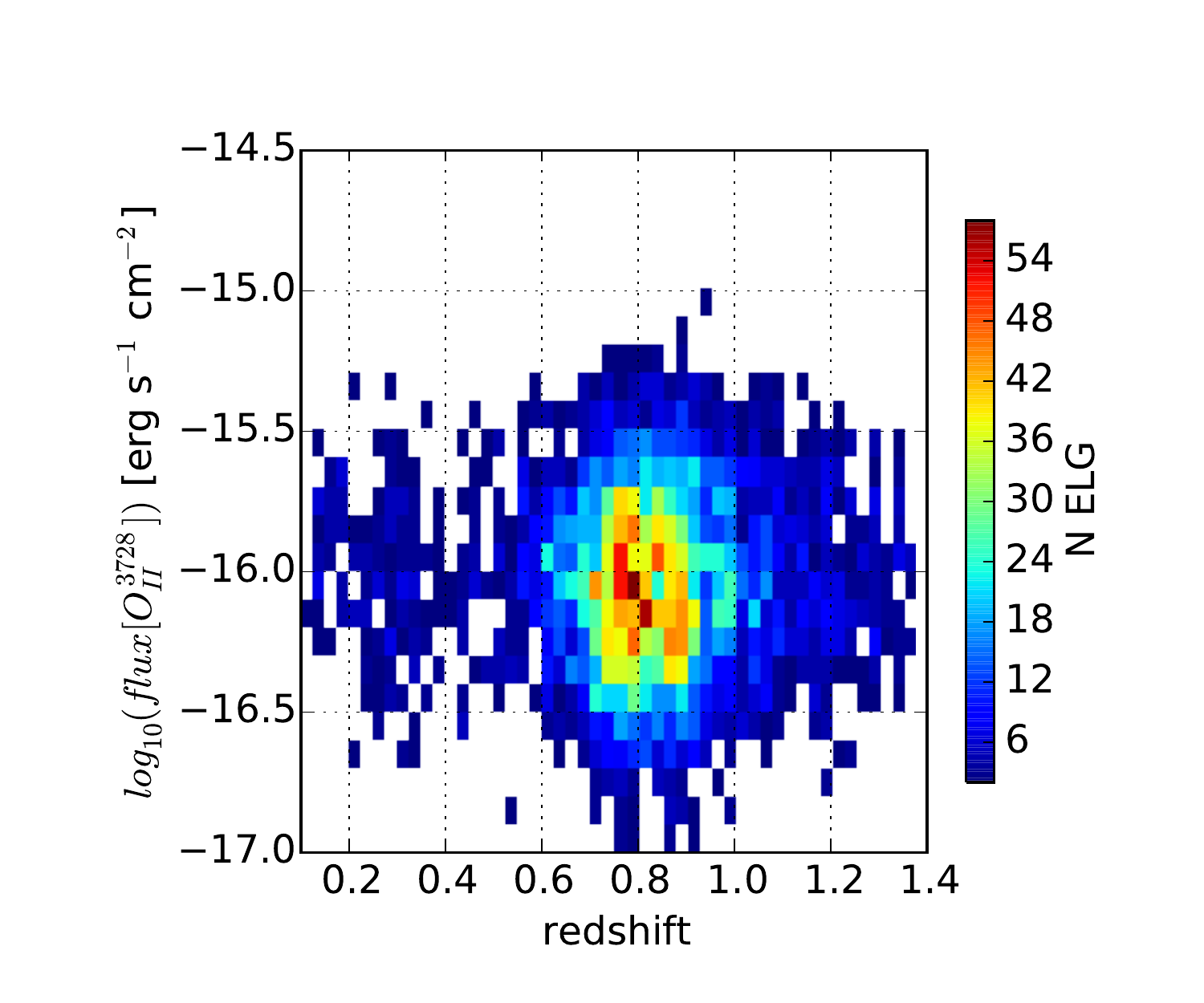}
\caption{Summary of eBOSS ELG pilot observations. RA vs. DEC colored with Galactic extinction (top left). The area covered has a low Galactic extinction. Number of ELGs as a function of redshift and \OIII, \Hb, and \OII line flux when measured with a signal-to-noise ratio higher than 5. \OII is the strongest emission line and is seen throughout the redshift range.}
\label{fig:all:DATA:BOSS}
\end{center}
\end{figure*}

\begin{figure*}
\begin{center}
\includegraphics[height=0.4\textwidth]{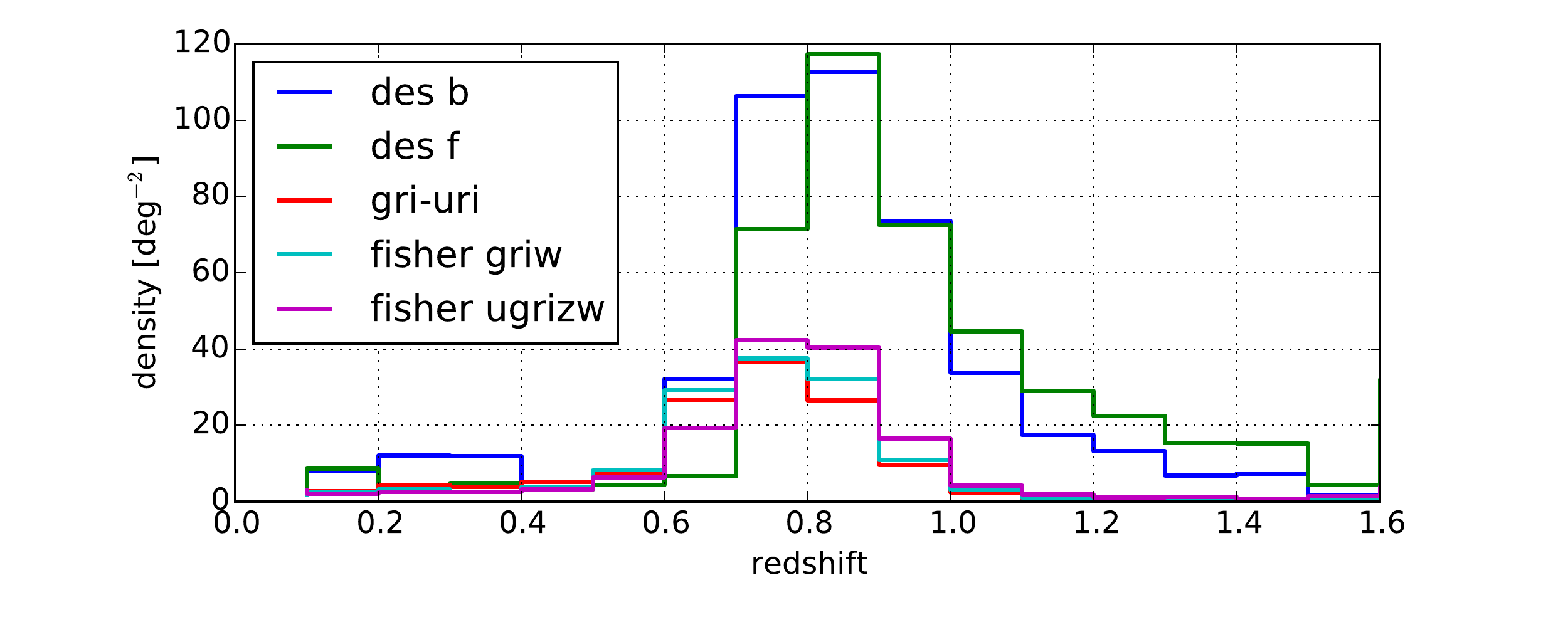}
\caption{Redshift distribution obtained for each selection. The DES-based selections have a higher density and a higher mean redshift. The identification rates (galaxies with redshifts) are des-b: 76\%, des-f: 71\%, gri-uri: 68\%, fisher griw: 72\%, and fisher ugrizw: 76\%.}
\label{fig:all:DATA:BOSS:NZ}
\end{center}
\end{figure*}

Using the photometric catalogs described previously and the target selection algorithms described in this section, we targeted 9000 ELGs (chunks eboss 6 and 7). The positions of the observed targets on the sky and the measured line flux as a function of redshift for the main emission lines \OII, \Hb, and \OIII are presented in Fig. \ref{fig:all:DATA:BOSS}. The position of the observations on the sky are shown in the first panel, coded by the level of Galactic extinction \citep{1998ApJ...500..525S}: we observed a region with low extinction. 
The other three panels show that most lines have a flux greater than $\log$(flux / \uflux)$>-16.5$. 
In this section, we detail how each selection was designed and describe their efficiency in targeting ELGs in the redshift range 0.6 to 1.1. We use the criterion to automatically select reliable redshifts given in Eq. (\ref{zOK:selection}).

\subsection{Observational strategy}
For this test, we designed seven different selection schemes. Two of them are based on the DES photometry, the others on the SDSS-WISE-SCUSS photometry. We gave first priority to the selections based on the SDSS-WISE-SCUSS photometry because they have lower surface densities and are limited by brighter magnitude cuts. We gave second priority to the fainter and denser DES-based selections. Table \ref{eBOSS:ELG:observation:summary:table} summarizes all the selections probed, their density, and the observed fraction. As a result, the first-priority selections had more than 90\% of their targets assigned fibers and the second-priority selections had between 60 and 70\% of their targets observed. 

For the first-priority objects, the targets are drawn randomly and the obtained sample is a fair subsample of the parent sample. For the second-priority objects, we apply conditional weights to the second-priority objects to compute moments of its distribution or to produce histograms because there are some common targets in the
second- and first-priority selections.

The observations were designed with large overlaps between the plates to neglect fiber collision in the central area $34<RA<38^\circ$ and $-6<dec<-3.8$ where \citet{2015arXiv150907121J} computed the clustering of the observed ELGs.

We designed a broad selection to then re-cut the final selection for eBOSS inside these observations. This is important to have some data around the fiducial selection to understand the effects of each cut on the observed galaxy population. These observations constitute a super-set of the actual eBOSS ELG sample in color-magnitude redshift space.

\subsection{ELG selection with SDSS-WISE-SCUSS}
\label{selectionDetailsSSW}
Using the photometry from SDSS-WISE-SCUSS, we tested two approaches, a color selection and a Fisher linear discriminant \citep{AHG:AHG2137} selection, attempting to maximize both the share of ELGs in the redshift range of interest and the mean redshift of the sample with a lowest density of 180 ELG per deg$^2$.
Selecting ELGs at this redshift with such a photometry slightly pushes the limits. There is only little wiggle room for the magnitude limit, which drags the density of tracers to the lower limit.

The priority scheme did not influence the completeness from one sample to another, and the observed samples are fair subsamples of the complete selections. 

\subsubsection{Color selection}
The color selection uses two color spaces $U-r-i$ and $g-r-i$ \citep[see][who defined the filters]{1996AJ....111.1748F}. We call this selection the gri-Uri selection. This is a further optimization of the color selections observed in \citet{2013MNRAS.428.1498C}.
It selects a mean of 197 targets per deg$^2$ (averaged over $\sim$50 deg$^2$) by applying the following selections on the photometric catalog. 
\begin{enumerate}
\item RESOLVE\_STATUS in the SDSS photometry has SURVEY\_PRIMARY on
\item $g_{\rm model}, r_{\rm model}, i_{\rm model}>$ 0
and $ g^{\rm err}_{\rm model}<$ 0.6
and $ r^{\rm err}_{\rm model}<$ 1
and $ i^{\rm err}_{\rm model}<$ 0.4  
\item and [ (a) OR (b) ] where
\begin{enumerate}
\item 
21 $\leq g_{\rm model} <22.5$
and $r_{\rm model}<$ 22.5
and $i_{\rm model}<$ 21.6
and $g_{\rm model} - r_{\rm model}<$ 0.8
and $r_{\rm model} - i_{\rm model}>$ 0.8  
\item  
20$<g_{\rm model}<$ 23
and $r_{\rm model}<$ 22.5
and $i_{\rm model}<$ 21.6
and $21<U_{\rm modelAdd}<22.5$
and $r_{\rm model} - i_{\rm model}>$ 0.7
and $r_{\rm model} - U_{\rm modelAdd}> 0.7 - 3.5*(r_{\rm model} - i_{\rm model}), $  
\end{enumerate}
\end{enumerate}
where magnitudes are dust-extinction corrected using the coefficients from \citet{1998ApJ...500..525S}. 
$mag_{\rm model}$ are from the model magnitudes from SDSS DR12 \citep{2015ApJS..219...12A}, $U_{\rm modelAdd}$ is the $U$-band model magnitude from SCUSS. Given the uncertainty on the magnitude at the depth of the selection ($g<22.5$ and $U<22.5$), all coefficients in the selection were rounded to the first decimal without affecting the properties of the selected galaxy population.
A total of 2,484 targets were observed in a 13.36 deg$^2$ region, which corresponds to a target sampling rate TSR=N$_{\rm observed}$/N$_{\rm targeted}$= 94.4\%. The TSR does not depend on magnitude or on redshift, and the observations have enough overlap so that fiber collision is negligible. The observed sample is thus a fair sample of the complete sample. In Table \ref{table:nz:ELGtest:A} we report the redshift distributions and estimate the uncertainty on the number density by approximating the distribution per bin to follow a Poisson distribution, that is, that the uncertainty on N is $\sigma_N=N\sqrt{N_{obs}}/N_{obs}$.
We securely measured the redshifts of 68$\pm$2\% of the targets as galaxies or quasars; see Table \ref{table:nz:ELGtest:A}. This sample has a mean redshift of 0.734. The percentage of detections classified as stars is 7.7\%, which leaves a 25\% fraction of unknown objects.

\subsubsection{Fisher selections}
The Fisher selection algorithm allows us to make additionally
optimized color-selections in a greater number of dimensions, which slightly improves the selection efficiency.  For a thorough description of the Fisher selection method, see \citet{2016A&A...585A..50R}.
We constituted a training spectroscopic data sample using the SDSS, BOSS ELG, VIPERS, DEEP2, zCOSMOS, and VVDS surveys \citep[][respectively]{2015ApJS..219...12A,2015A&A...575A..40C,2014A&A...566A.108G,2013ApJS..208....5N,Lilly_2009,2013A&A...559A..14L} to derive the best possible selection with a Fisher discriminant. We test this selection function in the chunk eboss7, that is, a sub-region of the eboss6 chunk.
We consider two types of selection, first using the combination of SDSS and WISE, designated griW,  and second, using the combination of SCUSS, SDSS, and WISE, named UgrizW.
We emphasize that the selections published in \citet{2016A&A...585A..50R} are slightly different from those tested in eboss7 because the selection function was further optimized based on the eboss6-7 tests. The redshift distributions obtained using the Fisher approach are given in Table \ref{table:nz:ELGtest:B}.

We use the first two filters of the WISE photometry W1 and W2 to construct a composite extinction-corrected AB magnitude, $W_{\rm model}$, which takes advantage of the two measurements \citep[see][for the details about the WISE composite magnitude]{2015ApJS..221...27M}. First, we convert W1 and W2 magnitudes to the AB magnitude system (AB = Vega + dm, with dm(W1)=2.699 and dm(W2)=3.339). We then define W as follows:
\begin{enumerate}
\item $W$ = W1, if no W2 measurement
\item $W$ = W2, if no W1 measurement
\item else: flux($W$) = (flux(W1) + 0.5*flux(W2))/1.5 when both W1 and W2 measurements are present. 
\end{enumerate}
For the uncertainty, we construct $W^{\rm err}_{\rm model}$ from W1\_MAGERR and W2\_MAGERR using the propagation of uncertainties in quadrature. For the extinction, we consider extinction\_W = extinction\_W1 because abs(extinction\_W1-extinction\_W2)$<$5e-3 in the test region). Finally, we use $W_{\rm model}=W-extinction\_W1$.

\subsubsection*{Fisher UgrizW}

The Fisher UgrizW selection selection was adopted as a first priority and contains
\begin{enumerate}
\item   RESOLVE\_STATUS in the SDSS photometry has SURVEY\_PRIMARY on
\item  OBJC\_TYPE=3 or $r_{\rm model}>$22
\item  20.0$<g_{\rm model}<$22.7 and $g^{\rm err}_{\rm model}<$0.5
and 19.0$<r_{\rm model}<$22.5 and $r^{\rm err}_{\rm model}<$0.5
and 19.0$<i_{\rm model}<$21.5 and $i^{\rm err}_{\rm model}<$0.5
and 17.0$<W_{\rm model}<$21.0 and $W^{\rm err}_{\rm model}<$0.5
and $U_{\rm modelAdd}>0$
\item  1.23$<$Fisher$_{\rm UgrizW}<$5.0, 
\end{enumerate}
with Fisher$_{\rm UgrizW}= - 0.390197(u_{\rm modelAdd} - r_{\rm model}) - 0.497885(g_{\rm model} - r_{\rm model}) + 0.0734933(r_{\rm model} - i_{\rm model}) + 0.480957(r_{\rm model} - W_{\rm model}) + 0.152151(r_{\rm model} - z_{\rm model}) + 0.847598$.
It has a target density of 199 per square degree. A total of 76\% are identified as galaxies or QSOs with a mean redshift at 0.788. The detections classified as stars are 1.7\%, and 22\% remain unknown.

\subsubsection*{Fisher UgrizW bright}

The Fisher selection UgrizW bright selection was chosen as a second priority; the overlap with the first selection is broad.
\begin{enumerate}
\item Same as the Fisher UgrizW selection points 1 to 3 (not 4)
\item $g_{\rm model}<$22.5 and 1.13$<$Fisher$_{\rm UgrizW}<$1.23. 
\end{enumerate}
It has a target density of 43 per square degree; 71\% are identified as galaxies or QSOs with a mean redshift at 0.724. The detections classified as stars are 0.34\%, and 28\% remain unknown. It seems less efficient, but the uncertainty on the efficiency is 9\% because there are only a few targets.

\subsubsection*{Fisher UgrizW bright all}

The Fisher selection UgrizW bright all is the same as the UgrizW bright, but without the FisherUgrizW$<$1.23 selection.
It has a target density of 196 per square degree; 78\% are identified as galaxies or QSOs with a mean redshift at 0.778. The detections classified as stars are 1.47\%, and 20\% remain unknown.

\subsubsection*{Fisher griW}

The Fisher selection griW was selected as a third priority and does not use the SCUSS U band or the SDSS $z$ band, which are quite shallow.
\begin{enumerate}
\item    RESOLVE\_STATUS in the SDSS photometry has SURVEY\_PRIMARY on
\item   OBJC\_TYPE=3 or $r_{\rm model}>$22
\item   20.0$<g_{\rm model}<$22.5 and $g^{\rm err}_{\rm model}<$0.5
and 19.0$<r_{\rm model}<$22.5 and $r^{\rm err}_{\rm model}<$0.5
and 19.0$<i_{\rm model}<$21.5 and $i^{\rm err}_{\rm model}<$0.5
and 17.0$<W_{\rm model}<$21.0 and $W^{\rm err}_{\rm model}<$0.5
\item   0.61$<$Fisher$_{\rm griW}<$5.00, 
\end{enumerate}
with Fisher$_{\rm griW}= -0.50972(g_{\rm model}-r_{\rm model}) +0.304366(r_{\rm model}-i_{\rm model}) +0.353073(r_{\rm model}-W_{\rm model}) +0.0306172 $. 
This selection has a target density of 196 per square degree; 72\% are identified as galaxies or QSOs with a mean redshift at 0.788. The detections classified as stars are 2.61\%, and 25\% remain unknown.

Table \ref{table:nz:ELGtest:B} gives the detailed density of galaxies observed as a function of redshift using the Fisher approach.

\subsubsection{Results with the SDSS-WISE-SCUSS selections}

For the same target density, the Fisher selections improve the identification rate by up to 10\%, increase the mean redshift, and diminish the contamination by lower redshift galaxies and stars compared to the color selection. The first decile of the redshift distribution is higher than 0.6 compared to 0.43 for the gri-Uri selection, see Table \ref{eBOSS:ELG:observation:summary:NZs}.

The Fisher selection was further optimized by \citet{2016A&A...585A..50R}, and it meets the requirement.

\begin{table*}
\caption{ELG selection target surface densities (after applying the bright star mask) and observed surface densities. The first five selections are based on SDSS-WISE-SCUSS and the last two on DES. For each selection scheme the first line gives the result of the eboss6 chunk and the second line the combination of eboss6 and eboss7 chunks. Note that the areal extent of eboss7 is contained in eboss6.}
\begin{center}
\begin{tabular}{l rrr rrr r}
\hline\hline
selection & \multicolumn{3}{|c|}{Targeted}  & \multicolumn{3}{|c|}{Observed} & fraction\\ 
name & N & area & density  & N & area & density &  observed \\ 
& & [deg$^{2}$] & [deg$^{-2}$] && [deg$^{2}$] & [deg$^{-2}$] & [\%]\\ \hline
gri-Uri  & 9686  &  49.18  &  196.94  &  2484  &  13.36  &  185.90  &  94.39  \\ 
-  & - &  -  & - &  1588  &  8.82  &  180.11  &  91.45  \\ \hline
Fisher griW  & 9639  &  49.18  &  195.99  &  1375  &  13.36  &  102.90  &  52.50  \\ 
-  & - &  -  & - &  1621  &  8.82  &  183.85  &  93.81  \\ \hline
Fisher UgrizW bright  & 2143  &  49.18  &  43.57  &  188  &  13.36  &  14.07  &  32.29  \\ 
-   & - &  -  & - &  303  &  8.82  &  34.37  &  78.87  \\ \hline
Fisher UgrizW bright all  & 9676  &  49.18  &  196.74  &  1287  &  13.36  &  96.32  &  48.96  \\ 
-  & - &  -  & - &  1595  &  8.82  &  180.90  &  91.95  \\ \hline
Fisher UgrizW  & 9798  &  49.18  &  199.22  &  1204  &  13.36  &  90.10  &  45.23  \\ 
-  & - &  -  & - &  1696  &  8.82  &  192.36  &  96.55  \\ \hline
des bright  & 8306  &  13.50  &  615.26  &  3272  &  9.20  &  355.65  &  57.81  \\ 
-  & - &  -  & - &  3842  &  8.82  &  435.75  &  70.82  \\ \hline
des faint  & 8776  &  13.50  &  650.07  &  3249  &  9.20  &  353.15  &  54.32  \\ 
-  & - &  -  & - &  3406  &  8.82  &  386.30  &  59.42  \\ \hline
\end{tabular}
\end{center}
\label{eBOSS:ELG:observation:summary:table}
\end{table*}

\subsection{ELG selection with DECam - DES photometry}
\label{selectionDetailsDES}

We performed two tests selecting either brighter and redder galaxies (DES bright) or fainter and bluer galaxies (DES faint). These objects are targeted as last-priority objects after the gri-Uri targets in eboss6 and after the Fisher targets in eboss7. 

\subsubsection{DES, bright selection}
The bright selection is defined by
\begin{enumerate}
\item  20.5 $< g_{\rm model} <$ 22.8
\item and -0.7 $<g_{\rm model} - r_{\rm model} <$ 0.9
and 0 $<r_{\rm model} -z_{\rm model} <$ 2
and $r_{\rm model} -z_{\rm model} > 0.4*(g_{\rm model} -r_{\rm model} ) + 0.4$
\item $g_{\rm 2"} - g_{\rm model}<$ 2
and $r_{\rm 2"} - r_{\rm model}<$ 2
and $z_{\rm 2"} - z_{\rm model}<$ 2: it rejects false detections near bright stars or saturated bright stars,
\end{enumerate}
where $mag_{\rm model}$ is the model magnitude (MAG\_DETMODEL) and $mag_{\rm 2"}$ is the 2'' diameter aperture magnitude (MAG\_APER\_4) reported by the DES pipeline. We designate this selection by des bright.
It has a target density of 615 per square degree; 76\% are identified as galaxies with a mean redshift at 0.843. The detections classified as stars are 7.7\%, and 16\% remain unknown.

\subsubsection{DES, faint selection}
The faint selection is defined by
\begin{enumerate}
\item $g_{\rm model}>$ 20.45 and $r_{\rm model}<$ 22.79
\item 0.285 $<r_{\rm model}-z_{\rm model}<$ 1.585
 and $g_{\rm model} - r_{\rm model} $<$ 1.1458\, (r_{\rm model} - z_{\rm model}) - 0.209$
 and  $ g_{\rm model} - r_{\rm model} $<$ 1.4551 - 1.1458 \, (r_{\rm model} - z_{\rm model})$
\item $g_{\rm 2"} - g_{\rm model}<$ 2
and $r_{\rm 2"} - r_{\rm model}<$ 2
and $z_{\rm 2"} - z_{\rm model}<$ 2.
\end{enumerate}
It has a target density of 650 per square degree. Seventy-one
percent are identified as galaxies with a mean redshift at 0.9. The detections classified as stars are 6.3\% and 23\% remain unknown. The observed redshift distribution is given in Table \ref{table:nz:ELGtest:C}.  We designate this selection by des faint.

\subsubsection{Results with DECam - DES photometry}

The DES-based selections were observed at a lower completeness level (60\%-70\%) and the average redshift is higher than with the SDSS-based selections. We weight the observations accordingly to recover the correct redshift distribution. The redshift distributions are too extended for the purpose of eBOSS.

\subsection{Optimized selection using DECam imaging}
\label{sec:decamTS}
Using the eboss6 and 7 observations, we further optimize the DECam-based target selection to increase its efficiency for the purpose of eBOSS ELG selection. The further optimization of the Fisher algorithms is presented in \citet{2016A&A...585A..50R}. The eBOSS ELG program will target 255,000 ELGs. We will draw targets from 1,500 deg$^2$ of imaging observed by DES or DECaLS. The target density of fibers assigned to ELGs is therefore 170 deg$^{-2}$. Given a 95\% fiber assignment efficiency, we need to provide 180 deg$^{-2}$.

\begin{figure}
\begin{center}
\includegraphics[height=80mm]{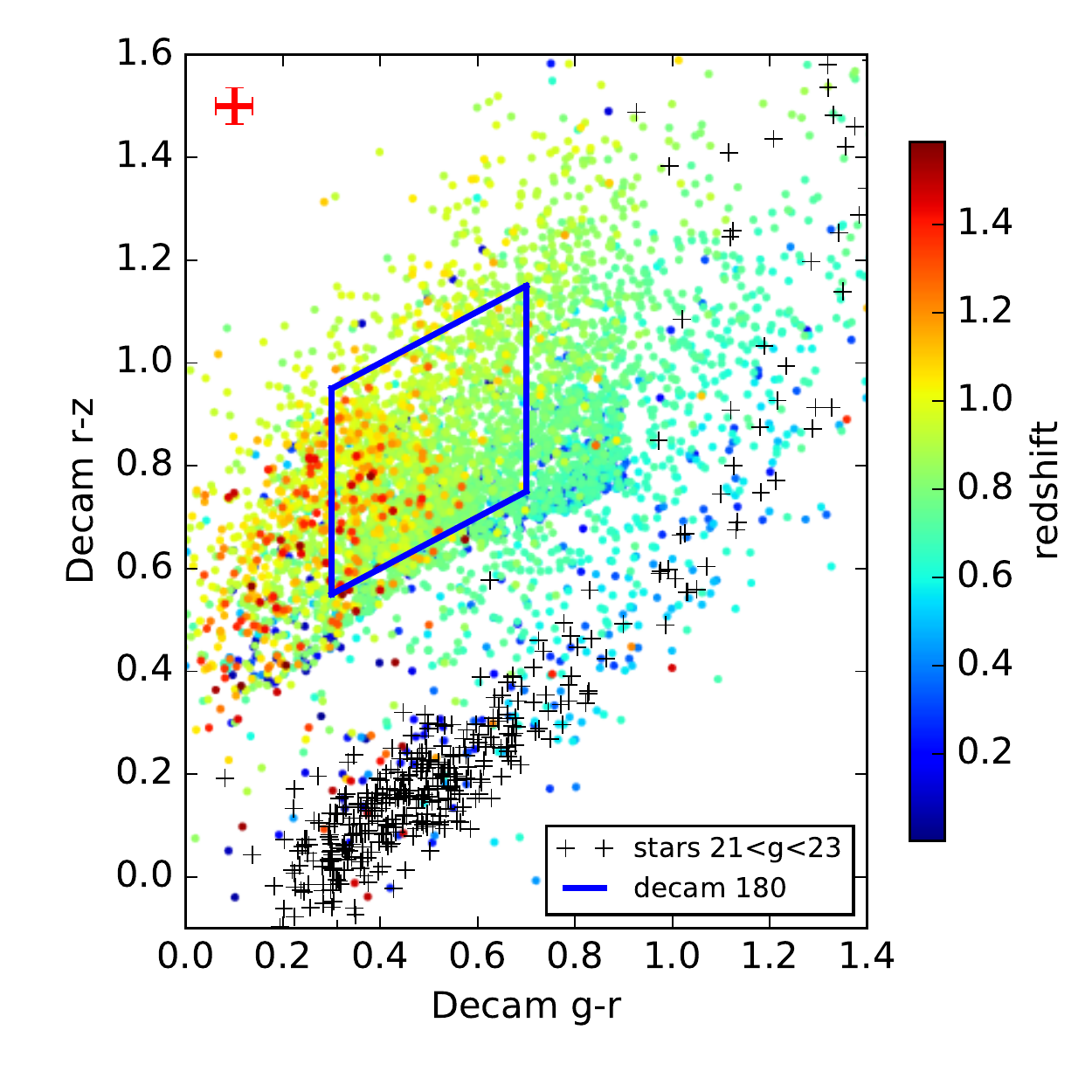}
\caption{Color (g-r) -color (r-z) diagram showing the optimized DECam-based ELG selection algorithms. We show all the good spectroscopic redshifts observed by eboss6 and 7. The mean error on the DES colors is shown with the red cross in the top left corner.}
\label{fig:TS:ELG:DES}
\end{center}
\end{figure}

We found a further optimization of the DES-based ELG algorithms that satisfies the eBOSS requirements. We name this algorithm decam 180. It has the same bright-star-contamination exclusion scheme: 
$g_{\rm 2"} - g_{\rm model}<$ 2
and $r_{\rm 2"} - r_{\rm model}<$ 2
and $z_{\rm 2"} - z_{\rm model}<$ 2. 
Then, the ELG selection is
\begin{enumerate}
\item $22.1< g_{\rm model} < 22.8$
\item $0.3< g_{\rm model} -r_{\rm model} < 0.7$
and $0.25 < r_{\rm model} -z_{\rm model}  < 1.4$
\item $0.5(g_{\rm model} -r_{\rm model}) + 0.4< r_{\rm model} -z_{\rm model}  <  0.5(g_{\rm model} -r_{\rm model}) + 0.8$.
\end{enumerate} 
 Figure \ref{fig:TS:ELG:DES} shows the selection box in the $g-r$, $r-z$ plane.
 
It selects 182$\pm3$ targets per deg$^2$ and outputs $N(0.7<z_{good}<1.1)=75.8\pm0.4\%$ with a mean redshift at 0.864. The redshift quality flags discard 4\% of the total amount of redshifts in the range $0.7<z<1.1$. These are detections with low S/N, and one half of the redshift are incorrect and one half is correct, which means that future pipeline improvement could increase the efficiency by 2\% at
most.
The redshift identification rate is 87$\pm0.5\%$. The contaminants are low-redshift galaxies (2.3\%), higher redshift galaxies (7.7\%), and unknown redshifts (14\%). The median \OII (\OIII) flux observed is  7 (6) $\times 10{-17}$\uflux. These numbers are based on the observation of 2,604 (decam 180) targets in eboss6 and eboss7. For the decam 180 selection, fewer than 0.5\% of the targets fall in the category SEL.

The redshift distribution is given in Tables \ref{eBOSS:ELG:observation:summary:NZs} and \ref{table:NZ:desSelections}. It has a narrow redshift range and is mainly contaminated by low-quality redshifts and higher redshift, $z>1.1$, galaxies.
An independent study of this selection on the COSMOS \OII catalog \citep{2015A&A...575A..40C} and its corresponding DECam photometry\footnote{\url{http://legacysurvey.org/}} provides the same redshift distributions and success rates.

The selection scheme presented was chosen from a handful of other selections producing similar efficiencies and densities. To do so, we sorted all possible selection schemes with the mean \OII emission line flux so that this scheme guarantee strong lines.

Using the current spectroscopic data (eboss6-7) and applying the same method as described in \citet{2015arXiv150907121J}, we measure the monopole clustering and deduce the galaxy bias for both samples: $b=1.7\pm0.1$. The further optimization of the selection increased the efficiency without changing the clustering amplitude. 

\section{Pilot observations, results}
We compare the target density and redshift distribution of 
the observed and optimized selection algorithms in Tables \ref{eBOSS:selection:summary} and \ref{eBOSS:ELG:observation:summary:NZs}. 
It is difficult to target ELGs beyond redshift one using the SDSS-WISE-SCUSS selections photometry, but it is feasible with the DES photometry. Beyond redshift 1.1, the \OII  doublet is not systematically split, and the efficiency in assigning correct redshift therefore diminishes. Sky lines also become stronger.
The optimized decam algorithm meets the requirements set by eBOSS (we note that the observed selections do not meet the requirements). In addition, the optimized Fisher is at the limit of meeting the requirements with a 71 \% efficiency.

Using shallower photometry to reach the density of galaxies required by eBOSS, we are forced to target near the limit of the photometry. In this regime, the Malmquist bias becomes non-negligible and the actual mean magnitude of the selection is fainter than the magnitude selection imposed on the data, so that the sample becomes dominated by fainter galaxies and the redshift identification rate decreases. 

\subsection{Insights on the object classification of plates 8123
and 8130}
We use the class information from the inspection made on the two eboss6 plates 8123 and 8130 to determine stellar contamination; see Table \ref{table:sdss:inspection}. The selection algorithms gri-Uri, des bright, and des faint mostly target galaxies. The contamination by quasars or stars is small. The main contamination is due to spectra with a low S/N where the redshift cannot be securely determined; these are probably faint galaxies. The current automated classification that identifies the type of object (star, galaxy, or quasar) is only reliable for high S/N spectra and not for the low S/N spectra where the class assignments should not be trusted.

\begin{table*}
\caption{Summary of the selections ordered by median redshift. The efficiency is the number density of galaxies identified in the redshift range specified divided by the total number density of galaxies. The column photometry lists the combination of photometry. 1: SCUSS + SDSS, 2: SCUSS + SDSS + WISE, 3: SDSS + WISE, and 4: DECam. The uncertainties on the number densities and efficiencies are comprised between 1.5 and 3 percent, but for the sake of readability, we report all uncertainties in the tables in the
Appendix. The last column reports whether the efficiency requirement are met (the density requirement is always met). }
\begin{center}
\begin{tabular}{l ccccccccc }
\hline\hline
selection & photometry & magnitude selection & density & median  & ID rate & efficiency  & efficiency &  \checkmark ? \\
name & & upper bound &  [deg$^{-2}$] & redshift & & $0.6<z<1$ & $0.7<z<1.1$\\
\hline
gri-Uri  & 1 & $g<22.5$ or $U<22.5$ & 197 & 0.734  & 0.68 & 0.52 & 0.40 &x \\ 
griW  & 2 & $g<22.5$ & 196 & 0.767 &  0.72  & 0.56 & 0.42 &x \\ 
UgrizW bright all & 3 & $g<22.5$ & 196 & 0.778 & 0.78 & 0.59 & 0.49 &x\\
UgrizW  & 3 & $g<22.7$ & 199 & 0.788 & 0.76 &  0.59 & 0.52 &x\\
des bright & 4 &$g<22.8$ & 615 & 0.843 & 0.76 & 0.53 & 0.53&x \\
des faint & 4 &$r<22.8$ & 650 &  0.901 & 0.71 & 0.41 & 0.47 &x \\
\hline
Fisher griW optimized & 2& $g<22.5$ & 182 & 0.76 & 0.85 & 0.71 & 0.54&  $\sim$\checkmark \\
decam 180 & 4& $g<22.8$ & 182 & 0.864 & 0.87 & 0.7 & 0.76&  \checkmark \\
\hline
\end{tabular}
\end{center}
\label{eBOSS:selection:summary}
\end{table*}

\begin{table}
\caption{Moments of the eBOSS ELG redshift distribution: first decile D10, first, second (median), and third quartiles Q25, Q50, Q75, and last decile D90.}
\begin{center}
\begin{tabular}{l cccccccc }
\hline\hline
selection &D10 & Q25 & Q50 & Q75 & D90 \\
\hline
gri-Uri  &0.432 & 0.633 & 0.734 & 0.820 & 0.904 \\ 
griW  &0.604 & 0.690 & 0.767 & 0.852 & 0.945 \\ 
UgrizW bright  &0.600 & 0.661 & 0.724 & 0.814 & 0.928 \\
UgrizW bright all  &0.628 & 0.710 & 0.778 & 0.862 & 0.956 \\
UgrizW  & 0.639 & 0.723 & 0.788 & 0.866 & 0.958 \\
des bright  &0.619 & 0.745 & 0.843 & 0.964 & 1.198 \\
des faint  &0.729 & 0.801 & 0.901 & 1.119 & 1.441 \\
decam 180 &0.754 & 0.794 & 0.864 & 0.932&  1.077 \\ \hline
\end{tabular}
\end{center}
\label{eBOSS:ELG:observation:summary:NZs}
\end{table}
\begin{table*}
\caption{Result of the inspection of two plates from eboss6. The last column is the identification rate of the spectral class by the inspectors.}
\begin{center}
\begin{tabular}{c r r r r r rr r }
\hline\hline
name & N & [ galaxy / quasar ]& star & low S/N & id rate [\%] \\ \hline
gri-Uri & 858  &  603  [  589  /  14  ] &  17  &  238  &  70.28  \\
des bright & 635  &  530  [  523  /  07  ] &  0  &  105  &  83.46  \\
des faint & 658  &  530  [  512  /  18  ] &  1  &  127  &  80.55  \\ \hline
\end{tabular}
\end{center}
\label{table:sdss:inspection}
\end{table*}


\section{Summary}
\label{sec:Discussion}
In this article, we demonstrated that the automated redshift estimation for ELGs at redshift 0.8 is reliable, in particular, we confined the possible line confusion rate to a sub-percent level. We also documented the eBOSS ELG pilot survey: 9,000 new spectra targeted from different photometric surveys. We provided reliable redshift distributions for each selection scheme. We additionally optimized and finalized one of the possible eBOSS ELG selections using DECam-based imaging. This selection has a density of 180 for a galaxy bias of 1.7$\pm0.1$ and an efficiency of nearly 76 percent. This selection is best suited for a wide-angle survey to precisely measure the BAO in the two-point correlation function at redshift 0.85.

\subsection*{Future plans}
The ELG samples under construction will be extremely useful for investigating the galaxy population that forms stars most efficiently.

The spectroscopic signature of ELGs is quite specific and mixes light emitted by the stellar population (which is a combination of recently formed and older stars and light reprocessed by the interstellar medium) and by the circumgalactic medium. To provide a global panchromatic view of these galaxies, we will study the infrared light emitted by their dust component in a future publication.

By combining $N$-body simulations with semi-analytical models that reproduce observations, we aim to develop the analysis reported in \citet{2015arXiv150704356F} to provide a more complete description of ELG properties, clarify their nature, and maximize their potential for constraining cosmological models.

\section*{Acknowledgements}
\vspace{0.2cm}
JC and FP acknowledge support from the Spanish MICINNs Consolider-Ingenio 2010 Programme under grant MultiDark CSD2009-00064, MINECO Centro de Excelencia Severo Ochoa Programme under the grants SEV-2012-0249, FPA2012-34694, and the projects AYA2014-60641-C2-1-P and AYA2012-31101.
We also thank the Lawrence Berkeley National Laboratory for its hospitality. FP acknowledges the spanish MEC Salvador de Madariaga program, Ref. PRX14/00444.
TD and JPK acknowledge support from the LIDA ERC advanced grant. AR acknowledges funding from the P2IO LabEx (ANR-10-LABX-0038) in the framework Investissements d'Avenir (ANR- 11-IDEX-0003-01) managed by the French National Research Agency (ANR). EJ acknowledges the support of CNRS and the Labex OCEVU.

This paper represents an effort by the SDSS-III, SDSS-IV and DES collaborations.

Funding for SDSS-III was provided by the Alfred
P. Sloan Foundation, the Participating Institutions, the
National Science Foundation, and the U.S. Department
of Energy Office of Science. The SDSS web site is www.sdss.org.

SDSS-IV acknowledges support and resources from the Center for High-Performance Computing at
the University of Utah. 

SDSS-IV is managed by the Astrophysical Research Consortium for the
Participating Institutions of the SDSS Collaboration including the
Brazilian Participation Group, the Carnegie Institution for Science,
Carnegie Mellon University, the Chilean Participation Group,
the French Participation Group, Harvard-Smithsonian Center for Astrophysics,
Instituto de Astrof\'isica de Canarias, The Johns Hopkins University,
Kavli Institute for the Physics and Mathematics of the Universe (IPMU) /
University of Tokyo, Lawrence Berkeley National Laboratory,
Leibniz Institut f\"ur Astrophysik Potsdam (AIP),
Max-Planck-Institut f\"ur Astronomie (MPIA Heidelberg),
Max-Planck-Institut f\"ur Astrophysik (MPA Garching),
Max-Planck-Institut f\"ur Extraterrestrische Physik (MPE),
National Astronomical Observatory of China, New Mexico State University,
New York University, University of Notre Dame,
Observat\'ario Nacional / MCTI, The Ohio State University,
Pennsylvania State University, Shanghai Astronomical Observatory,
United Kingdom Participation Group,
Universidad Nacional Aut\'onoma de M\'exico, University of Arizona,
University of Colorado Boulder, University of Portsmouth,
University of Utah, University of Virginia, University of Washington,
University of Wisconsin,
Vanderbilt University, Yale University and the french participation group.

Funding for the DES Projects has been provided by the U.S. Department of Energy, the U.S. National Science Foundation, the Ministry of Science and Education of Spain, 
the Science and Technology Facilities Council of the United Kingdom, the Higher Education Funding Council for England, the National Center for Supercomputing 
Applications at the University of Illinois at Urbana-Champaign, the Kavli Institute of Cosmological Physics at the University of Chicago, 
the Center for Cosmology and Astro-Particle Physics at the Ohio State University,
the Mitchell Institute for Fundamental Physics and Astronomy at Texas A\&M University, Financiadora de Estudos e Projetos, 
Funda{\c c}{\~a}o Carlos Chagas Filho de Amparo {\`a} Pesquisa do Estado do Rio de Janeiro, Conselho Nacional de Desenvolvimento Cient{\'i}fico e Tecnol{\'o}gico and 
the Minist{\'e}rio da Ci{\^e}ncia, Tecnologia e Inova{\c c}{\~a}o, the Deutsche Forschungsgemeinschaft and the Collaborating Institutions in the Dark Energy Survey. 

The Collaborating Institutions are Argonne National Laboratory, the University of California at Santa Cruz, the University of Cambridge, Centro de Investigaciones Energ{\'e}ticas, 
Medioambientales y Tecnol{\'o}gicas-Madrid, the University of Chicago, University College London, the DES-Brazil Consortium, the University of Edinburgh, 
the Eidgen{\"o}ssische Technische Hochschule (ETH) Z{\"u}rich, 
Fermi National Accelerator Laboratory, the University of Illinois at Urbana-Champaign, the Institut de Ci{\`e}ncies de l'Espai (IEEC/CSIC), 
the Institut de F{\'i}sica d'Altes Energies, Lawrence Berkeley National Laboratory, the Ludwig-Maximilians Universit{\"a}t M{\"u}nchen and the associated Excellence Cluster Universe, 
the University of Michigan, the National Optical Astronomy Observatory, the University of Nottingham, The Ohio State University, the University of Pennsylvania, the University of Portsmouth, 
SLAC National Accelerator Laboratory, Stanford University, the University of Sussex, and Texas A\&M University.

The DES data management system is supported by the National Science Foundation under Grant Number AST-1138766.
The DES participants from Spanish institutions are partially supported by MINECO under grants AYA2012-39559, ESP2013-48274, FPA2013-47986, and Centro de Excelencia Severo Ochoa SEV-2012-0234.
Research leading to these results has received funding from the European Research Council under the European Union's Seventh Framework Programme (FP7/2007-2013) including ERC grant agreements 
 240672, 291329, and 306478.
 
 We are grateful for the extraordinary contributions of our CTIO colleagues and the DECam Construction, Commissioning and Science Verification
teams in achieving the excellent instrument and telescope conditions that have made this work possible.  The success of this project also 
relies critically on the expertise and dedication of the DES Data Management group.

This paper includes targets derived from the images of
the Wide-Field Infrared Survey Explorer, which is a
joint project of the University of California, Los Angeles,
and the Jet Propulsion Laboratory/California Institute
of Technology, funded by the National Aeronautics and
Space Administration.

This paper has gone through internal review by the DES collaboration.

\bibliographystyle{aa}
\bibliography{ELG_observation}

\appendix
\section{Redshift distributions}

\begin{table}
\caption{Redshift distribution for reliably identified redshifts per square degree during eboss6 observations N$=N_{obs}(z_{min}<z\leq z_{max})/TSR/$area. The error given is taken from a Poisson distribution: $\sigma_N=N/\sqrt{N_{obs}}$. The area is 13.36 deg$^{2}$.}
\begin{center}
\begin{tabular}{cc r r }
\hline \hline
\multicolumn{2}{c}{redshift} & 
\multicolumn{2}{c}{gri-Uri} \\
$z_{min}$ & $z_{max}$ &
 N [deg$^{-2}$] & $\sigma_N$ \\ \hline
0.0  &  0.1  &  2.06  &  0.40  \\ 
0.1  &  0.2  &  2.62  &  0.46  \\ 
0.2  &  0.3  &  3.49  &  0.53  \\ 
0.3  &  0.4  &  3.96  &  0.56  \\ 
0.4  &  0.5  &  4.68  &  0.61  \\ 
0.5  &  0.6  &  9.12  &  0.85  \\ 
0.6  &  0.7  &  26.64  &  1.45  \\ 
0.7  &  0.8  &  41.39  &  1.81  \\ 
0.8  &  0.9  &  24.97  &  1.41  \\ 
0.9  &  1.0  &  10.47  &  0.91  \\ 
1.0  &  1.1  &  2.46  &  0.44  \\ 
1.1  &  1.2  &  0.40  &  0.18  \\ 
1.2  &  1.3  &  0.24  &  0.14  \\ 
1.3  &  1.4  &  0.48  &  0.19  \\ 
1.4  &  1.5  &  0.24  &  0.14  \\ 
1.5  &  1.6  &  0.16  &  0.11  \\ 
1.6  &  2.4  &  0.71  &  0.24  \\ 
\hline
\multicolumn{2}{c}{total}& 
134.07  &  3.25  \\ 
\multicolumn{2}{c}{ID rate }&  
0.68  &    \\ 
   \hline
\end{tabular}
\end{center}
\label{table:nz:ELGtest:A}
\end{table}

\begin{table*}
\caption{Same as Table \ref{table:nz:ELGtest:A} for the Fisher selections observed by eboss6 or eboss7 on the eboss7 footprint: the area is 8.82 deg$^{2}$.}
\begin{center}
\begin{tabular}{rrrrrrrrrrrrrrrrrrrr}
\hline \hline
\multicolumn{2}{c}{redshift} & 
 \multicolumn{2}{c}{UgrizW} & 
 \multicolumn{2}{c}{UgrizW bright} & 
 \multicolumn{2}{c}{UgrizW bright all} & 
 \multicolumn{2}{c}{griW} \\
$z_{min}$ & $z_{max}$ &
 N & $\sigma_N$ &  N & $\sigma_N$ & 
 N & $\sigma_N$ &  N & $\sigma_N$  \\ \hline
0.0  &  0.1  &  2.94  &  0.59  &  0.74  &  0.30  &  3.56  &  0.65  &  1.90  &  0.47  \\ 
0.1  &  0.2  &  2.00  &  0.48  &  0.61  &  0.27  &  2.49  &  0.54  &  2.38  &  0.53  \\ 
0.2  &  0.3  &  2.47  &  0.54  &  0.36  &  0.21  &  2.71  &  0.57  &  3.21  &  0.62  \\ 
0.3  &  0.4  &  2.58  &  0.55  &  0.25  &  0.17  &  2.48  &  0.54  &  2.62  &  0.56  \\ 
0.4  &  0.5  &  3.17  &  0.61  &  1.23  &  0.39  &  3.93  &  0.68  &  3.80  &  0.67  \\ 
0.5  &  0.6  &  6.34  &  0.86  &  2.45  &  0.55  &  7.15  &  0.92  &  8.22  &  0.99  \\ 
0.6  &  0.7  &  19.38  &  1.51  &  8.41  &  1.01  &  23.21  &  1.66  &  29.26  &  1.87  \\ 
0.7  &  0.8  &  42.29  &  2.23  &  8.33  &  1.01  &  41.46  &  2.22  &  37.60  &  2.11  \\ 
0.8  &  0.9  &  40.41  &  2.18  &  4.41  &  0.74  &  36.48  &  2.08  &  32.11  &  1.95  \\ 
0.9  &  1.0  &  16.45  &  1.39  &  1.59  &  0.44  &  14.63  &  1.31  &  10.86  &  1.13  \\ 
1.0  &  1.1  &  4.23  &  0.70  &  0.98  &  0.35  &  4.98  &  0.77  &  3.06  &  0.60  \\ 
1.1  &  1.2  &  1.88  &  0.47  &  0.12  &  0.12  &  1.77  &  0.46  &  1.06  &  0.35  \\ 
1.2  &  1.3  &  1.06  &  0.35  &  0.25  &  0.17  &  1.30  &  0.39  &  0.47  &  0.24  \\ 
1.3  &  1.4  &  1.29  &  0.39  &  0.12  &  0.12  &  1.18  &  0.37  &  0.71  &  0.29  \\ 
1.4  &  1.5  &  0.59  &  0.26  &  0.37  &  0.21  &  0.96  &  0.34  &  0.59  &  0.26  \\ 
1.5  &  1.6  &  1.41  &  0.41  &  0.24  &  0.17  &  1.42  &  0.41  &  0.94  &  0.33  \\ 
1.6  &  2.4  &  3.88  &  0.67  &  0.62  &  0.28  &  4.26  &  0.71  &  2.95  &  0.59  \\ 
\hline
\multicolumn{2}{c}{total} &  152.35  &  4.22  &  31.09  &  1.95  &  153.96  &  4.27  &  141.76  &  4.09  \\ 
\multicolumn{2}{c}{ID rate}& 0.76  & &  0.71  & &  0.78  & &  0.72  \\ 
\hline
  \end{tabular}
\end{center}
\label{table:nz:ELGtest:B}
\end{table*}

\begin{table}
\caption{Same as Table \ref{table:nz:ELGtest:A} for the DES selections, based on eboss6-7 spectra in the eboss7 area that has a higher completeness.}
\begin{center}
\begin{tabular}{rrrrrrrrrr}
\hline \hline
\multicolumn{2}{c}{redshift} & 
\multicolumn{2}{c}{des-b} & 
\multicolumn{2}{c}{des-f}  \\
$z_{min}$ & $z_{max}$ & 
 N & $\sigma_N$ &  N & $\sigma_N$  \\ \hline
0.0  &  0.1  &  1.77  &  0.51  &  4.06  &  0.83  \\ 
0.1  &  0.2  &  8.21  &  1.14  &  8.64  &  1.21  \\ 
0.2  &  0.3  &  12.09  &  1.39  &  3.92  &  0.82  \\ 
0.3  &  0.4  &  11.97  &  1.38  &  4.84  &  0.90  \\ 
0.4  &  0.5  &  3.37  &  0.70  &  3.14  &  0.72  \\ 
0.5  &  0.6  &  7.79  &  1.08  &  4.39  &  0.83  \\ 
0.6  &  0.7  &  32.18  &  2.19  &  6.63  &  1.02  \\ 
0.7  &  0.8  &  106.24  &  3.95  &  71.49  &  3.44  \\ 
0.8  &  0.9  &  112.60  &  4.05  &  117.35  &  4.39  \\ 
0.9  &  1.0  &  73.51  &  3.32  &  72.50  &  3.44  \\ 
1.0  &  1.1  &  33.83  &  2.29  &  44.64  &  2.71  \\ 
1.1  &  1.2  &  17.44  &  1.65  &  28.94  &  2.19  \\ 
1.2  &  1.3  &  13.20  &  1.44  &  22.51  &  1.94  \\ 
1.3  &  1.4  &  6.80  &  1.04  &  15.33  &  1.60  \\ 
1.4  &  1.5  &  7.28  &  1.07  &  15.27  &  1.60  \\ 
1.5  &  1.6  &  1.52  &  0.48  &  4.37  &  0.86  \\ 
1.6  &  2.4  &  18.35  &  1.69  &  32.01  &  2.29  \\ 
 \hline
\multicolumn{2}{c}{total} & 468.15  &  8.34& 460.05  &  8.64  \\ 
\multicolumn{2}{c}{ID rate}&0.76  & &  0.71  \\ 
\hline
\end{tabular}
\end{center}
\label{table:nz:ELGtest:C}
\end{table}

\begin{table}
\caption{Redshift distribution for the DECam-optimized algorithm selecting $\sim$182 ELG per deg$^2$.}
\begin{center}
\begin{tabular}{r r r r r r r r r }
\hline \hline
 \multicolumn{2}{c}{redshift} &  \multicolumn{2}{c}{decam 180}  \\
  \multicolumn{1}{c}{$z_{min}$} &
  \multicolumn{1}{c}{$z_{max}$} &
  N & $\sigma_N$ &\\  \hline
0.00&0.10&0.29&0.01   \\ 
0.10&0.20&0.87&0.07   \\ 
0.20&0.30&0.76&-0.20   \\ 
0.30&0.40&1.49&0.21   \\ 
0.40&0.50&0.35&0.03   \\ 
0.50&0.60&0.70&0.06   \\ 
0.60&0.70&1.78&0.22   \\ 
0.70&0.80&35.26&0.53   \\ 
0.80&0.90&61.37&1.50   \\ 
0.90&1.00&28.78&0.31   \\ 
1.00&1.10&13.19&1.07   \\ 
1.10&1.20&6.05&0.53   \\ 
1.20&1.30&4.51&0.19   \\ 
1.30&1.40&1.28&-0.16   \\ 
1.40&1.50&1.12&0.00   \\ 
1.50&1.60&0.00&0.00   \\ 
1.60&1.70&0.00&0.00   \\ 
1.70&1.80&0.00&0.00   \\ 
1.80&1.90&0.00&0.00   \\ 
1.90&2.00&0.15&0.04   \\ 
2.00&2.10&0.15&0.04   \\ 
2.10&2.20&0.25&-0.07   \\ 
2.20&2.30&0.00&0.00   \\ 
2.30&2.40&0.17&0.01   \\ \hline
\multicolumn{2}{c}{total [deg$^{-2}$]} & 158.5  &  -   \\ 
\multicolumn{2}{c}{ID rate [\%]}& 0.86  & -  \\ 
\hline\end{tabular}
\end{center}
\label{table:NZ:desSelections}
\end{table}

\end{document}